\documentclass[aps,twocolumn,superscriptaddress,showpacs,showkeys]{revtex4}
\usepackage{graphics,graphicx,dcolumn,bm,fleqn,epic,eepic,float}
\usepackage{amssymb,amsmath,multirow,rotate,color,float,times}
\usepackage{color}
\usepackage{soul}                    
\definecolor{red}{rgb}{1,0,0}
\definecolor{green}{rgb}{0,1,0}
\definecolor{blue}{rgb}{0,0,1}

\newcommand{\bite}{\begin{itemize}}
\newcommand{\eite}{\end{itemize}}
\newcommand{\benu}{\begin{enumerate}}
\newcommand{\eenu}{\end{enumerate}}
\newcommand{\bverb}{\begin{verbatim}}
\newcommand{\everb}{\end{verbatim}}

\newcommand{\beq}{\begin{equation}}
\newcommand{\eeq}{\end{equation}}
\newcommand{\barr}{\begin{array}}
\newcommand{\earr}{\end{array}}
\newcommand{\p}{\partial}

\newcommand{\mbf}{\mathbf}
\newcommand{\bdif}{\,\boldsymbol{d}}
\newcommand{\fracpp}[1]{\frac{\p}{\p#1}}

\newcommand{\mom}[1]{m^{(#1)}}
\newcommand{\moms}[1]{m^{*(#1)}}
\newcommand{\h}[1]{h^{(#1)}}
\newcommand{\bmom}[1]{\mbf{m}^{(#1)}}
\newcommand{\bmoms}[1]{\mbf{m}^{*(#1)}}
\newcommand{\bh}[1]{\mbf{h}^{(#1)}}
\newcommand{\coef}[1]{c^{(#1)}}
\newcommand{\coefa}[1]{a^{(#1)}}
\newcommand{\coefb}[1]{b^{(#1)}}
\newcommand{\bcoefa}[1]{\mbf{a}^{(#1)}}
\newcommand{\bcoefb}[1]{\mbf{b}^{(#1)}}

\newcommand{\vecx}{\mathbf{x}}
\newcommand{\vecX}{\mathbf{X}}

\newcommand{\vecY}{\mathbf{Y}}
\newcommand{\vecz}{\mathbf{z}}

\newcommand{\vecomega}{\boldsymbol{\omega}}
\newcommand{\matA}{\mathbf{A}}
\newcommand{\matB}{\mathbf{B}}
\newcommand{\matC}{\mathbf{C}}
\newcommand{\matM}{\mathbf{M}}
\newcommand{\matQ}{\mathbf{Q}}
\newcommand{\matV}{\mathbf{V}}
\newcommand{\numax}{{\nu_\text{max}}}
\newcommand{\Four}{{\cal F}}
\newcommand{\Mon}{{\cal M}}

\newcommand{\DDrift}{D^{(1)}}
\newcommand{\DDiff}{D^{(2)}}
\newcommand{\vecDDrift}{\mbf{D}^{(1)}}
\newcommand{\vecDDiff}{\mbf{D}^{(2)}}

\newcommand{\Ito}{It\^o~}
\begin{document}

\title{Analysis of stochastic time series in N dimensions in the presence of strong measurement noise}

\author{B.~Lehle}
\affiliation{bernd@vflow.de, vFlow Engineering GmbH, Pforzheimer Strasse 348, D-70499 Stuttgart, Germany}

\begin{abstract}
An extension and generalization of a recently presented approach for the analysis of Langevin-type stochastic processes
in the presence of strong measurement noise is presented. For a stochastic process in $N$ dimensions which is
superimposed with strong, exponentially correlated, Gaussian distributed, measurement noise it is possible to
extract the strength and the correlation functions of the noise as well as polynomial approximations of the
drift and diffusion functions of the underlying process.
\end{abstract}

\pacs{05.40.Ca,  
      02.50.Ey}  

\keywords{Measurement noise, Stochastic processes}

\maketitle

\section{Introduction}
\label{sec:intro}

In the last years there has been significant progress in the analysis and characterization of the dynamics
of processes underlying the time series of complex dynamical systems \cite{friedrich11,friedrich08,schreiberbook,abarbanel}.
If the temporal evolution of a quantity $\vecX(t)$ can be described by a Langevin equation, it is possible to extract drift
and diffusion functions of the underlying stochastic process from a given timeseries.
This can be done because the moments of the conditional probability densities of $\vecX(t\!+\!\tau)|_{\vecX(t)=\vecx}$
can be related to these functions.

Since this approach was introduced \cite{friedrich97,ryskin97,siegert98,friedrich00,gradisek00}
it has been successfully carried out in a broad range of fields. For example for data from financial markets
\cite{finance00}, traffic flow \cite{trafic02}, chaotic electrical circuits \cite{circuit03,circuit04}, human heart beat
\cite{heart04}, climate indices \cite{climate04,climate05}, turbulent fluid dynamics \cite{fluid07}, and for
electroencephalographic data from epilepsy patients \cite{epilepsy08,epilepsy09}.

Real-world data, however, also gives rise to some problems. One of them is, that experimental data is only given with a
finite sampling rate. So methods had to be proposed to deal with the effects arising from this
fact \cite{sampling02,sampling05,sampling08,honisch11,sampling12MLE}.
Another problem is the virtually unavoidable measurement noise \cite{schreiberbook,noise93,noise00,sampling08}.
In the presence of measurement noise $\vecY(t)$ the values of $\vecX(t)$ or any of its probability densities are
no longer accessible, but only $\vecX^*(t)=\vecX(t)+\vecY(t)$ and {\em its} density distributions.

Recently an approach has been presented which allows the estimation of drift and diffusion functions
in the presence of strong, delta-correlated, Gaussian noise \cite{iterative06,iterative10}.
Starting with initial estimates for the noise strength and the drift and diffusion functions a functional
of these unknowns is iteratively minimized. An alternative approach, which is able to deal also with strong, exponentially
correlated, Gaussian noise, has been presented in \cite{lehle11}.

The aim of this paper is the formulation of this later approach in $N$ dimensions and also its generalization. The basic
idea stays the same. Instead of looking at the conditional moments in the first place, the joint probability density
$\rho(\vecx,\vecx'\!,\tau)$ of pairs $(\vecX(t),\vecX(t\!+\!\tau))$ is looked at.
If the measurement noise is independent of $\vecX(t)$, then $(\vecX(t),\vecX(t\!+\!\tau))$ and
$(\vecY(t),\vecY(t\!+\!\tau))$ are independent random variables. Hence the joint probability density
$\rho^*(\vecx,\vecx'\!,\tau)$ of their sum $(\vecX^*(t),\vecX^*(t\!+\!\tau)$ is given by the convolution of $\rho$ and $\rho_Y$, where
$\rho_Y(\vecx,\vecx'\!,\tau)$ is the joint probability density of $(\vecY(t),\vecY(t\!+\!\tau))$.

The noise is assumed to be Gaussian and the Gauss function has some special algebraic properties. This allows to
express the moments of $\rho^*$ in terms of the moments of $\rho$ and of the noise parameters. The obtained relations
can then be used to extract the noise parameters. Furthermore, by the use of integral transformes (the Fourier
transform used in \cite{lehle11} is a special case hereof), it is possible to extract polynomial approximations of
the drift and diffusion functions using purely algebraic relations between quantities that can be calculated directly
from a given, noisy time series.

This paper is organized as follows: Section \ref{sec:process} is devoted to the noise-free stochastic process, the
definition of its joint probability density and expressions for the moments of this density in terms of a Taylor-It\^o
expansion. Section \ref{sec:noise} provides the properties of the measurement noise under consideration and in section
\ref{sec:noisy_process} expressions for the moments of a noisy process will be derived. After looking at the benefits
of equidistantly sampled experimental time series in section \ref{sec:assumption series}, the previously derived
expressions will be used in section \ref{sec:extracting noise} to extract the parameters of the measurement noise
and in section \ref{sec:extracting coeffs} to extract polynomial approximations for drift and diffusion functions.
Finally in section \ref{sec:examples} the results will be applied to some synthetic time series. The used properties 
of the Gauss function and further computational details are given in appendices \ref{app:gauss} and \ref{app:cond_mom}.

\section{Stochastic process}
\label{sec:process}

Let $\vecX(t)$ be a stochastic process in $N$ dimensions that can be described by a time-independent \Ito-Langevin
equation

\begin{eqnarray}
d\vecX_i(t) &=& \vecDDrift(\vecX)\,dt+\sqrt{\vecDDiff(\vecX)}\; d\mbf{W}(t) \label{langevin},
\end{eqnarray}

\noindent where $\vecDDrift$ and $\vecDDiff$ are the Kramers-Moyal coefficients of the corresponding Fokker-Planck
equation and $d\mbf{W}$ denotes a vector of increments of independent Wiener processes with $<\!dW_idW_j\!>=\delta_{ij}dt$.
The notation $\sqrt{\vecDDiff}$ is used to denote a matrix $\mbf{g}$ with $\mbf{g}\cdot\mbf{g}^t=\vecDDiff$ \cite{risken89}.

Let the one- and two-point probability density functions of $\vecX$ be denoted by

\begin{subequations}
\begin{eqnarray}
\rho(\vecx) &:=& p(\vecx,t)\\
\rho(\vecx,\vecx'\!,\tau) &:=& p(\vecx,t;\vecx'\!,t+\!\tau) \\
&=& \rho(\vecx)\,p(\vecx'\!,t+\!\tau|\vecx,t) \label{probdens1}
\end{eqnarray}
\end{subequations}

\noindent and let the conditioned moments of $\rho(\vecx,\vecx'\!,\!\tau)$ be defined as follows (the notation
$\bdif x$ is used to denote the product $dx_1\ldots dx_N$ whereas $d\vecx$ denotes a vector of differentials $dx_i$).

\begin{subequations}
\begin{eqnarray}
\mom{0}(\vecx) &=&\int_{\vecx'}\!\rho(\vecx,\vecx'\!,\!\tau)\bdif x'\\
\mom{1}_i(\vecx,\tau) &=&\int_{\vecx'}\!(x'_i\!-\!x_i)\rho(\vecx,\vecx'\!,\!\tau)\bdif x'\\
\mom{2}_{ij}(\vecx,\tau) &=&\int_{\vecx'}\!(x'_i\!-\!x_i)(x'_j\!-\!x_j)\rho(\vecx,\vecx'\!,\!\tau)\bdif x'\!
\end{eqnarray}
\end{subequations}

\noindent These moments are observable quantities. For a given time series they can be estimated by binning or
other density-estimation techniques. Using Eq.~(\ref{probdens1}) allows to express the moments $\bmom{k}$ in terms of
moments $\bh{k}$ of the conditional increments of $\vecX$

\begin{subequations}
\begin{eqnarray}
\mom{0}(\vecx) &=&\rho(\vecx)\cdot 1\\
\mom{1}_i(\vecx,\tau) &=&\rho(\vecx)\cdot\h{1}_i(\vecx,\tau)\\
\mom{2}_{ij}(\vecx,\tau) &=&\rho(\vecx)\cdot\h{2}_{ij}(\vecx,\tau),
\end{eqnarray}\label{def_moments}
\end{subequations}

\noindent with

\begin{subequations}
\begin{eqnarray}
\h{1}_i(\vecx,\tau) &:=& <\![X_i(t+\!\tau)-X_i(t)]\Big|_{\vecX(t)=\vecx\!}> \\
\h{2}_{ij}(\vecx,\tau) &:=& <\![X_i(t+\!\tau)-X_i(t)]\cr
                       && \times [X_j(t+\!\tau)-X_j(t)]\Big|_{\vecX(t)=\vecx}\!>.
\end{eqnarray}
\end{subequations}

\noindent A Taylor-\Ito expansion of  Eq.~(\ref{langevin}) provides expressions for these expectation values. Provided
that $\vecDDrift$ and $\vecDDiff$ are smooth functions in $\vecx$, it is possible to represent $\bh{1}$
and $\bh{2}$ as power series in $\tau$. The lowest order terms in these series are linear in $\tau$. The
series-coefficients are given by sums of products of the Kramers-Moyal coefficients and their derivatives and are thus
generally functions of $\vecx$.

\begin{subequations}
\begin{eqnarray}
\h{1}_i(\vecx,\tau) &=& \sum_{k=1}^\infty \coef{1,k}_i(\vecx)\,\tau^k\label{coeffs_h1}\\
\h{2}_{ij}(\vecx,\tau) &=& \sum_{k=1}^\infty \coef{2,k}_{ij}(\vecx)\,\tau^k
\end{eqnarray}\label{def_h1h2}
\end{subequations}

\noindent The explicit terms up to second order are given below (using index notation and summation convention). A
detailed description of the Taylor-\Ito expansion and its moments can be found in \cite{platen99}.

\begin{subequations}
\begin{eqnarray}
\h{1}_i    &=& \tau\DDrift_i\!+\!\frac{\tau^2}{2}\!\left[\DDrift_j\p_j\DDrift_i\right.\cr
            && \left.\!+\frac{1}{2}\DDiff_{jk}\p_j\p_k\DDrift_i\right]\!+\!O(\tau^3)\\
\h{2}_{ij} &=& \tau\DDiff_{ij}\!+\!\frac{\tau^2}{2}\!\left[2\DDrift_i\DDrift_j\!+\!\DDiff_{ik}\p_k\DDrift_j\right.\cr
           && \left.\!+\!\DDiff_{jk}\p_k\DDrift_i\!+\!\DDrift_k\p_k\DDiff_{ij}\right.\cr
           && \left.\!+\!\frac{1}{2}\DDiff_{kl}\p_k\p_l\DDiff_{ij}\right]\!+\!O(\tau^3)
\end{eqnarray}\label{condmoms_h1h2}
\end{subequations}

\noindent Inserting the series representations into Eq.~(\ref{def_moments}) yields a relation between
the observable moments $\bmom{k}$ and the unknown functions $\vecDDrift$ and $\vecDDiff$. For small values of $\tau$
this allows the direct estimation of the Kramers-Moyal coefficients.

\begin{subequations}
\begin{eqnarray}
\DDrift_i(\vecx) &=& \frac{1}{\tau}\frac{\mom{1}_i(\vecx,\tau)}{\mom{0}(\vecx)}+O(\tau)\\
\DDiff_{ij}(\vecx) &=& \frac{1}{\tau}\frac{\mom{2}_{ij}(\vecx,\tau)}{\mom{0}(\vecx)}+O(\tau)
\end{eqnarray}
\end{subequations}

\section{Measurement noise}
\label{sec:noise}

The measurement noise under consideration is denoted by $\vecY(t)$ and described by an Ornstein-Uhlenbeck process in $N$
dimensions. Such a process is characterized by linear drift- and constant diffusion functions and its statistical
properties can be derived analytically (see e.g. \cite{risken89}). The temporal evolution of $\vecY$ is described
by Eq.~(\ref{OUprocess}). Here the eigenvalues of matrix $\matA$ are required to have positive real part and matrix
$\matB$ is assumed to be symmetric and positive semi-definite.
The notation $\sqrt{\matB}$ is used to denote a matrix $\mbf{g}$ with $\mbf{g}\cdot\mbf{g}^t=\matB$ and the elements
of $d\mbf{W}$ denote the increments of independent Wiener processes with $<\!dW_idW_j\!>=\delta_{ij}dt$.

\begin{eqnarray}
d\vecY(t) &=& -\matA\cdot\vecY\, dt+\sqrt{\matB}\; d\mbf{W}(t).\label{OUprocess}
\end{eqnarray}

\noindent While $\matA$ and $\matB$ are appropriate to describe the temporal evolution of $\vecY$, the 'macroscopic'
properties of the noise are more conveniently described in terms of the covariance matrix $\matV$ and the
matrix of (exponentially decaying) correlation functions $\matM(\tau)$. Looking at the auto-covariance of $\vecY$ one finds

\begin{eqnarray}
<\!\vecY(t\!+\!\tau)\vecY^t(t)\!> &=& \matM(\tau)\matV
\end{eqnarray}

\noindent with

\begin{eqnarray}
\matM(\tau) &=& e^{-\matA \tau} \label{definition_M}\\
\matV &=& \int_0^\infty\! e^{-\matA s}\matB e^{-\matA^t s} ds.
\end{eqnarray}

\noindent Furthermore $\vecY$ is found to be Gaussian distributed. If $G(\matV,\vecx)$ is used to denote a normalized
Gauss function in $\vecx$ with covariance $\matV$ (see Eq.~(\ref{normGauss})), then the one- and two-point probability
density functions $\rho_Y$ can be written as

\begin{subequations}
\begin{eqnarray}
\rho_Y(\vecx) &=& G(\matV,\vecx)\\
\rho_Y(\vecx,\vecx'\!,\tau) &=& G(\matV,\vecx)\,G(\matC(\tau),\vecx'\!\!-\!\matM(\tau)\vecx) \label{pdf_noise}
\end{eqnarray}
\end{subequations}

\noindent with the shortcut

\begin{eqnarray}
\matC(\tau) &:=& \matV-\matM(\tau)\matV\matM^t(\tau). \label{def_matrixC}
\end{eqnarray}

\noindent A note on the eigenvalues of $\matA$ and $\matM$: If $\lambda_i$ denotes the eigenvalues of $\matA$, then the
eigenvalues of $\matM(\tau)$ are given by $e^{-\lambda_i\tau}$. By introducing the relaxation times (or characteristic
time scales) $T_i$ as $T_i:=1/\lambda_i$, the eigenvalues of $\matM$ can be written as $e^{-\tau/T_i}$. In the
numerical examples given later, the measurement noise will be characterized by such relaxation times $T_i$ instead of
by eigenvalues of $\matA$.

\section{Noisy stochastic process}
\label{sec:noisy_process}

Let $\vecX^*(t)=\vecX(t)+\vecY(t)$ be the sum of a stochastic signal $\vecX(t)$ and measurement noise $\vecY(t)$ as
introduced in sections~\ref{sec:process} and \ref{sec:noise}, respectively. Because $\vecX$ and $\vecY$ are
independent stochastic variables, the probability density functions of their sum $\vecX^*$ is given by the
convolution of the individual density functions.

\begin{eqnarray}
\rho^*(\vecx,\vecx'\!,\tau) &=& \rho_Y(\vecx,\vecx'\!,\tau)*\rho(\vecx,\vecx'\!,\tau)\cr
                &=& \int_{\vecz}\int_{\vecz'\!}\rho_Y(\vecx-\vecz,\vecx'\!-\vecz'\!,\tau)\cr
                &&  \times\quad \rho(\vecz,\vecz'\!,\tau)\bdif z\bdif z' \label{pdf_noisy}
\end{eqnarray}

\noindent Instead of the conditioned moments $\bmom{k}$ only their noisy counterparts $\bmoms{k}$ can be determined.

\begin{subequations}
\begin{eqnarray}
\moms{0}(\vecx) &=&\int_{\vecx'}\!\rho^*(\vecx,\vecx'\!,\!\tau)\bdif x'\\
\moms{1}_i(\vecx,\tau) &=&\int_{\vecx'}\!(x'_i\!-\!x_i)\rho^*(\vecx,\vecx'\!,\!\tau)\bdif x'\\
\moms{2}_{ij}(\vecx,\tau) &=&\int_{\vecx'}\!(x'_i\!-\!x_i)(x'_j\!-\!x_j)\cr
                          && \times \rho^*(\vecx,\vecx'\!,\!\tau)\bdif x'\!.
\end{eqnarray}\label{def_condmom_noisy}
\end{subequations}

\noindent Inserting the definitions of $\rho^*$ and $\rho_Y$ (Eqs.~(\ref{pdf_noisy}) and (\ref{pdf_noise})
respectively) and interchanging the order of integration, the integration with respect to $\vecx'\!$ can
be performed within the convolution integral. Using the definition of the moments $\bmom{k}$ and taking
advantage of the properties of the Gauss function then finally leads to the following equations (see
appendix~\ref{app:cond_mom}). Function arguments are omitted for notational simplicity.

\begin{subequations}
\begin{eqnarray}
\moms{0} &=&\rho_Y*\mom{0}\\
\moms{1}_i &=&\rho_Y*(\h{1}_i\mom{0}) + Q_{ii'}\p_{i'}\moms{0} \label{noisy_m1}\\
\moms{2}_{ij} &=&\rho_Y*(\h{2}_{ij}\mom{0}) \cr
              && + (Q_{ij}+Q_{ji}-Q_{ii'}Q_{jj'}\,\p_{i'}\p_{j'})\,\moms{0} \cr
              && + Q_{ii'}\p_{i'}\moms{1}_j + Q_{jj'}\p_{j'}\moms{1}_i \label{noisy_m2}
\end{eqnarray} \label{noisy_moments}
\end{subequations}

\noindent Here $\rho_Y(\vecx)=G(\matV,\vecx)$ is the density function of the measurement noise $\vecY$.
The terms $\bh{k}$, as introduced in section~\ref{sec:process}, denote the moments of the conditional
increments of $\vecX$. The quantity $\matQ$, finally, has been introduced as an abbreviation and is defined as

\begin{eqnarray}
\matQ(\tau) &:=& \bigl(\mbf{Id}-\matM(\tau)\bigr)\matV.
\end{eqnarray}

\noindent Equation (\ref{noisy_moments}) allows to express the observable moments $\bmoms{k}$ in terms of the
unknowns $\mom{0}$, $\bh{k}$, $\matM$ and $\matV$. However it is possible to use Eq.~(\ref{noisy_m1}) to extract the
parameters of the measurement noise without the need for a simultaneous determination of $\bh{k}$ and $\mom{0}$.
This will be done in section \ref{sec:extracting noise}. Next, however, an assumption on the given time series
will be made.

\section{Experimental time series}
\label{sec:assumption series}

It will be assumed, that the values of a given time series are taken at equidistant points in time with a basic time
increment of $\Delta t$. This is often assumed tacitly but shall be stated here explicitly because it will be used
in the following.

\begin{eqnarray}
\vecX^*_i &:=& \vecX^*(i\Delta t),\qquad i=1,\ldots,i_\text{max}
\end{eqnarray}

\noindent Increments of $\vecX^*$ can thus be calculated for all time increments $\tau$ which are integral
multiples of $\Delta t$. The experimental two-point probability density $\tilde\rho^*(\vecx,\vecx',\tau)$ for those
values of $\tau$ can then be written as a sum of Dirac-distributions.

\begin{eqnarray}
\tilde\rho^*(\vecx,\vecx',k\Delta t) &=& \frac{1}{i_\text{max}-k} \sum_{i=1}^{i_\text{max}-k}
                                    \delta(\vecx-\vecX^*_i)\cr
                               && \times\quad\delta(\vecx'-\vecX^*_{i+k})
\end{eqnarray}

\noindent Weighted integrals of the 'true' density $\rho^*$ can easily be estimated by weighted integrals of
$\tilde\rho^*$ now. Given a weight function $f(\vecx,\vecx')$ and denoting the estimate by $\tilde  I$ one finds

\begin{eqnarray}
\tilde I &=& \int_\vecx\int_{\vecx'} f(\vecx,\vecx')\tilde\rho^*(\vecx,\vecx',k\Delta t)\bdif x\bdif x'\cr
         &=& \frac{1}{i_\text{max}-k} \sum_{i=1}^{i_\text{max}-k}f(\vecX^*_i,\vecX^*_{i+k}).
\end{eqnarray}

\noindent Weighted integrals of $\rho^*$ can therefore directly be estimated from the given time series. There is no
need to use binning to estimate the density $\rho^*$ first \cite{lehle11}. Weighted integrals of the moments $\bmoms{k}$
can be treated the same way by expressing them as weighted integrals of $\rho^*$ using Eq.~(\ref{def_condmom_noisy}).

\section{Extracting measurement noise parameters}
\label{sec:extracting noise}

Multiplying Eq.~(\ref{noisy_m1}) by $x_j$ and subsequently applying an integration with
respect to $\vecx$ leads to

\begin{eqnarray}
\int_\vecx\moms{1}_ix_j\bdif x &=&\int_\vecx[\rho_Y*(\h{1}_i\mom{0})]x_j\bdif x\cr
             && + Q_{ii'}\int_\vecx(\p_{i'}\moms{0})x_j\bdif x. \label{extract_noise_1}
\end{eqnarray}

\noindent The left hand side of this equation can directly be estimated from a given time series and will
be denoted by $\mbf{Z}$.

\begin{eqnarray}
Z_{ij}(\tau) &:=& \int_\vecx\moms{1}_i(\vecx,\tau)\,x_j\bdif x
\end{eqnarray}

\noindent Applying integration by parts allows the evaluation of the second integral on the right hand side

\begin{eqnarray}
Q_{ii'}\int_\vecx(\p_{i'}\moms{0})x_j\bdif x &=& -Q_{ij}.
\end{eqnarray}

\noindent The remaining integral in Eq.~(\ref{extract_noise_1}) only depends on $\tau$ because of the function
$\bh{1}(\vecx,\tau)$. Using Eq.~(\ref{coeffs_h1}) therefore allows to express the integral as a power series in
$\tau$ with unknown coefficients $P^{(\nu)}_{ij}$. Truncating this series to some order $\numax$ yields
an approximation of the integral by a polynomial in $\tau$.

\begin{eqnarray}
\int_\vecx[\rho_Y*(\h{1}_i\mom{0})]x_j\bdif x &=& \sum_{\nu=1}^\numax P^{(\nu)}_{ij}\tau^\nu \label{def_poly_P}
\end{eqnarray}

\noindent Putting this all together (and additionally replacing the abbreviation $\matQ$ by its definition)
so far yields

\begin{eqnarray}
\mbf{Z}(\tau) &=& \sum_{\nu=1}^\numax \mbf{P}^{(\nu)}\tau^\nu-\bigl(\mbf{Id}-\matM(\tau)\bigr)\matV.
\end{eqnarray}

\noindent Assuming that the time series is sampled with a basic time increment $\Delta t$ (as stated
in section \ref{sec:assumption series}), the value of $\mbf{Z}(\tau)$ can directly be estimated for all
increments $\tau$ being integral multiples of $\Delta t$. The corresponding value of $\matM$ is
given by an integral power of $\matM(\Delta t)$ then. This is due to the fact that $\matM(\tau)$ (according
to section \ref{sec:noise}) is a matrix exponential. 

\begin{subequations}
\begin{eqnarray}
\matM_0 &:=& \matM(\Delta t) \;=\; e^{-\matA \Delta t}\\ 
\Rightarrow\quad\matM(k\Delta t) &=& e^{-\matA k\Delta t} \;=\; \matM_0^k 
\end{eqnarray}
\end{subequations}

\noindent So finally one gets

\begin{eqnarray}
\mbf{Z}(k\Delta t) &=& \sum_{\nu=1}^\numax \mbf{P}^{(\nu)}(k\Delta t)^\nu-\bigl(\mbf{Id}-\matM_0^k\bigr)\matV.
\end{eqnarray}

\noindent Evaluating $\mbf{Z}$ for a sufficient number of increments, $k\Delta t$,
yields a system of equations that can (iteratively) be solved for the unknowns $\mbf{P}^{(\nu)}$, $\matV$ and
$\matM_0$ in a least square sense. Subsequently the relaxation times of the measurement
noise, $T_i$, can be calculated from the eigenvalues of $\matM_0$.

For such a fit to succeed, two conditions must be met.
\bite
\item Firstly, the largest increment $k_\text{max}\Delta t$
should be small compared to the characteristic time scale of the underlying stochastic process. This will
allow, to chose a low polynomial order $\numax$ (the smaller $\tau$ the better $\bh{1}$ can be approximated by
low order polynomials).
\item Secondly, the relaxation times $T_i$ should be small compared to $k_\text{max}\Delta t$. This will allow, to
distinguish the exponential functions in $\matM(\tau)$ from a low order polynomial.
\eite

\noindent The proposed method is therefore limited to measurement noise with
relaxation times $T_i$ considerably smaller than the time scale of the underlying stochastic process.

\section{Extracting drift- and diffusion functions}
\label{sec:extracting coeffs}

In the following it will be assumed, that the noise parameters have already been estimated according to
section \ref{sec:extracting noise}. The parameters and derived quantities like $\matQ(\tau)$ will therefore be treated
as known quantities.

Multiplying Eqs.~(\ref{noisy_m1}) and (\ref{noisy_m2}) by some weight function $\Psi(\vecx)$ and subsequently applying
an integration with respect to $\vecx$ yields

\begin{subequations}
\begin{eqnarray}
\text{lhs}_i &=& \int_\vecx\Psi\left[\rho_Y*(\h{1}_i\mom{0})\right]\bdif x\label{weighted_m1}\\
\text{lhs}_{ij} &=& \int_\vecx\Psi\left[\rho_Y*(\h{2}_{ij}\mom{0})\right]\bdif x\label{weighted_m2}
\end{eqnarray}\label{weighted_m1m2}
\end{subequations}

\noindent where lhs$_i$ and lhs$_{ij}$ are abbreviations for the left hand sides

\begin{subequations}
\begin{eqnarray}
\text{lhs}_i &:=&\int_\vecx\Psi\left[\moms{1}_i-Q_{ii'}\p_{i'}\moms{0}\right]\bdif x\\
\text{lhs}_{ij} &:=&\int_\vecx\Psi\left[\moms{2}_{ij} - (Q_{ij}+Q_{ji}\right.\cr
                 && -Q_{ii'}Q_{jj'}\,\p_{i'}\p_{j'})\,\moms{0} \cr
                 && \left.- Q_{ii'}\p_{i'}\moms{1}_j - Q_{jj'}\p_{j'}\moms{1}_i\right]\bdif x.
\end{eqnarray}\label{def_lhs}
\end{subequations}

\noindent Applying integration by parts allows to express the integrals in Eq.~(\ref{def_lhs}) as sums of
weighted integrals of $\bmoms{k}$. For example one finds $\int\Psi\moms{1}_i+Q_{ii'}\int(\p_{i'}\Psi)\moms{0}$ for the
left hand side of Eq.~(\ref{weighted_m1}). This expression can directly be estimated from the given time series
because $\matQ$ and $\Psi$ (and thus also the derivatives of $\Psi$) are known. The same holds for the left hand
side of Eq.~(\ref{weighted_m2}). Both left hand sides can therefore directly be estimated for a
given choice of $\tau$ and $\Psi$.

The corresponding right hand sides, however, refer to the unknown function $\mom{0}$. It is possible to overcome
this problem if the drift- and diffusion functions are approximated by polynomials in $\vecx$.

\begin{subequations}
\begin{eqnarray}
\DDrift_i &=& \coefa{1}_{i}+\coefa{1}_{i\alpha}x_\alpha+\coefa{1}_{i\alpha\beta}x_\alpha x_\beta+\ldots\\
\DDiff_{ij} &=& \coefa{2}_{ij}+\coefa{2}_{ij\alpha}x_\alpha+\coefa{2}_{ij\alpha\beta}x_\alpha x_\beta+\ldots
\end{eqnarray}
\end{subequations}

\noindent The coefficients in the power series representation of the conditional moments $\bh{k}$ then also become
polynomials in $\vecx$.

\begin{subequations}
\begin{eqnarray}
\h{1}_i &=& \;\quad\,\tau\left[\coefa{1}_{i}+\coefa{1}_{i\alpha}x_\alpha+\coefa{1}_{i\alpha\beta}x_\alpha x_\beta+\ldots\right]\cr
      &&+\; \tau^2\left[\coefb{1}_{i}+\coefb{1}_{i\alpha}x_\alpha+\coefb{1}_{i\alpha\beta}x_\alpha x_\beta+\ldots\right]\cr
      &&+\;\ldots\\[0.5em]
\h{2}_{ij} &=& \;\quad\,\tau\left[\coefa{2}_{ij}+\coefa{2}_{ij\alpha}x_\alpha+\coefa{2}_{ij\alpha\beta}x_\alpha x_\beta+\ldots\right]\cr
      &&+\; \tau^2\left[\coefb{2}_{ij}+\coefb{2}_{ij\alpha}x_\alpha+\coefb{2}_{ij\alpha\beta}x_\alpha x_\beta+\ldots\right]\cr
      &&+\;\ldots
\end{eqnarray}\label{def_polyh1h2}
\end{subequations}

\noindent For the sake of simplicity the abbreviations $\bcoefb{k}$ have been introduced here. However all the
coefficients in Eq.~(\ref{def_polyh1h2}) can of cause be expressed in terms of the coefficients $\bcoefa{k}$.
Using Eq.~(\ref{def_polyh1h2}) the problem of expressing the right hand sides of Eq.~(\ref{weighted_m1m2}) reduces
to the problem of expressing terms of the form 

\begin{eqnarray}
F_{\alpha_1\ldots \alpha_k} &:=& \int_\vecx\Psi\left[\rho_Y*(x_{\alpha_1}\ldots x_{\alpha_k}\mom{0})\right]\bdif x.
\label{def_F_alpha}
\end{eqnarray}

\noindent Because $\rho_Y=G(\matV,\vecx)$ is a Gauss function, the convolution within the square brackets can be
expressed in terms of derivatives of $\moms{0}$, $x_\alpha\moms{0}$, $x_\alpha x_\beta\moms{0},\ldots$
(see appendix \ref{app:gauss:conv}). One finds

\begin{eqnarray}
\rho_Y*[\mom{0}] &=& \moms{0}\cr
\rho_Y*[x_\alpha\mom{0}] &=& x_\alpha \moms{0}+L_\alpha\moms{0}\cr
\rho_Y*[x_\alpha x_\beta\mom{0}] &=& x_\alpha x_\beta \moms{0}+L_\alpha[x_\beta \moms{0}]\cr
                                  &&+L_\beta[x_\alpha \moms{0}]+L_{\alpha\beta}\moms{0}\cr
                                 &\vdots&
\end{eqnarray}

\noindent with the linear differential operators

\begin{eqnarray}
L_\alpha &=& V_{\alpha\alpha'}\p_{\alpha'}\cr
L_{\alpha\beta} &=& V_{\alpha\alpha'}V_{\beta\beta'}\p_{\alpha'}\p_{\beta'}-V_{\alpha\beta}\cr
&\vdots&.
\end{eqnarray}

\noindent Applying integration by parts then allows to express Eq.~(\ref{def_F_alpha}) in terms of
weighted integrals of $\moms{0}$, which can directly be estimated from the given time series. One finds

\begin{eqnarray}
F &=& \int_\vecx\Psi \moms{0}\bdif x\cr
F_\alpha &=& \int_\vecx\left\{\Psi x_\alpha \moms{0}-[L_\alpha\Psi]\moms{0}\right\}\bdif x\cr
F_{\alpha\beta} &=& \int_\vecx\left\{\Psi x_\alpha x_\beta\moms{0}-[L_\alpha\Psi]x_\beta\moms{0}\right.\cr
 &&\left.-[L_\beta\Psi]x_\alpha\moms{0}+[L_{\alpha\beta}\Psi]\moms{0}\right\}\bdif x\cr
 &\vdots&.
\end{eqnarray}

\noindent Expressing the right hand sides of Eq.~(\ref{weighted_m1m2}) in terms of $F,F_\alpha,\ldots$ one
finally obtains the following equations. 

\begin{subequations}
\begin{eqnarray}
\text{lhs}_i &=& \;\quad\,\tau\left[\coefa{1}_{i}F+\coefa{1}_{i\alpha}F_\alpha+\coefa{1}_{i\alpha\beta}F_{\alpha\beta}+\ldots\right]\cr
      &&+\; \tau^2\left[\coefb{1}_{i}F+\coefb{1}_{i\alpha}F_\alpha+\coefb{1}_{i\alpha\beta}F_{\alpha\beta}+\ldots\right]\cr
      &&+\;\ldots\\[0.5em]
\text{lhs}_{ij} &=& \;\quad\,\tau\left[\coefa{2}_{ij}F+\coefa{2}_{ij\alpha}F_\alpha+\coefa{2}_{ij\alpha\beta}F_{\alpha\beta}+\ldots\right]\cr
      &&+\; \tau^2\left[\coefb{2}_{ij}F+\coefb{2}_{ij\alpha}F_\alpha+\coefb{2}_{ij\alpha\beta}F_{\alpha\beta}+\ldots\right]\cr
      &&+\;\ldots
\end{eqnarray}\label{coeff_fitting_final}
\end{subequations}

\noindent Evaluating the left hand sides and the quantities $F,F_\alpha,\ldots$ for a sufficient number of
increments $\tau$ and weight functions $\Psi$ yields a system of equations that can be solved for the unknown
polynomial coefficients in a least square sense. There are different approaches to deal with the higher order
terms in $\tau$ now.
\bite
\item The most simple approach is, to completely ignore the higher order terms. This will lead to a linear fit in $\tau$.
Furthermore the resulting set of equations will be linear in the unknown coefficients $\bcoefa{k}$.
\item A more elaborate approach is, to perform a polynomial fit in $\tau$. Coefficients beyond some order
will be ignored. If the remaining coefficients are treated as additional unknowns, then the resulting set of
equations will stay linear. However this way some available information is ignored, because $\bcoefb{k}$ and higher order
coefficients can in fact be expressed in terms of the coefficients $\bcoefa{k}$.
\item Finally, a polynomial fit in $\tau$ can be performed, where the higher order coefficients are expressed in terms
of the coefficients $\bcoefa{k}$. This will improve the accuracy of the estimate, because of the smaller number
of unknowns. The resulting set of equations, however, will now be nonlinear and needs to be solved iteratively.
\eite

\noindent The choice of the weight functions $\Psi$ has been left open up to now. Obviously $\Psi$ needs to admit
the various integrations by parts that have been applied. For these integrations it also has been tacitly assumed
that the involved boundary values at $|\vecx|\to\infty$ are vanishing. This imposes additional restrictions on $\Psi$.

The set of weight functions used in the numerical examples, given below, consisted of a number of Gauss functions
centered at different points in space. There may be better choices, but the problem of finding the 'best' set of
functions will not be addressed here. Gauss functions are smooth and real valued and have a local support. But maybe
it would be better to choose, for example, some complex valued functions like $\exp(i\vecomega^t\vecx)$ which have a
local support in Fourier space only. Also piecewise polynomial functions like the B-spline base functions may be an
alternative.

\section{Application to numerical data}
\label{sec:examples}

In order to check the accuracy of the proposed method, a test case in two dimensions has been investigated. A
stochastic process, $\vecX(t)$, as introduced in Sec.~\ref{sec:process}, has been specified by the following choice
for the drift- and diffusion functions ($x$ and $y$ denote the components of the vector $\vecx$).

\begin{subequations}
\begin{eqnarray}
\vecDDrift(\vecx) &=& \begin{pmatrix} x-xy \cr x^2-y \end{pmatrix}\\
\vecDDiff(\vecx) &=& \begin{pmatrix} 0.5 & 0 \cr 0 & 0.5(1+x^2) \end{pmatrix}
\end{eqnarray}
\end{subequations}

\begin{figure}[h]
\begin{center}
\includegraphics*[width=8.8cm,angle=0]{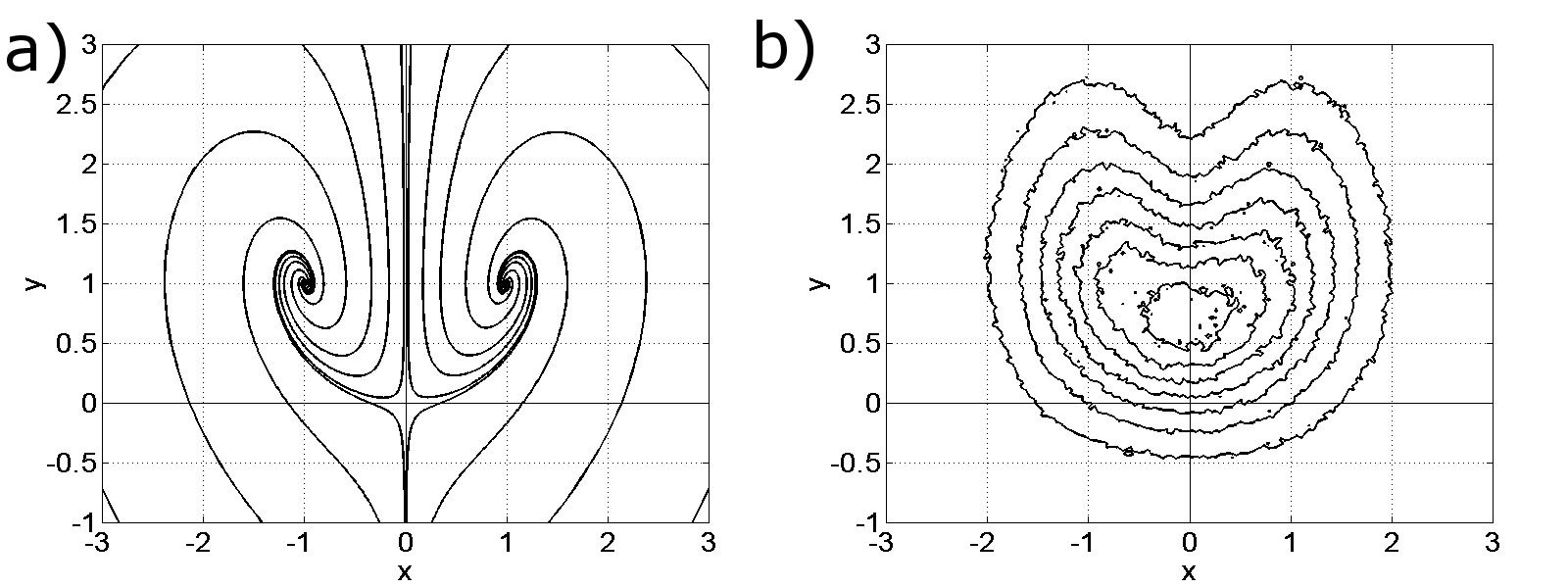}
\end{center}
\caption{\protect Deterministic dynamic and density distribution of the 2D process $\vecX(t)$.
The trajectories in phase space {\bf (a)} are generated by the deterministic part of the process dynamic ($\dot x=x-xy$,
$\dot y=x^2-y$). There exist three fixed points: a saddle at the origin and two stable foci at ($x=\pm 1,y=1$).
The contour lines {\bf (b)} of the probability density function have been computed from a time series of $\vecX$.
}
\label{fig_clean}
\end{figure}

\noindent By numerical integration synthetic time series of the process can be generated. All series used in the
following will consist of $10^7$ points, sampled at time increments $\Delta t=0.005$.
The deterministic part of the process dynamic and the experimental density distribution of $\vecX$ is visualized
in Fig.~\ref{fig_clean}.

\noindent The measurement noise, $\vecY(t)$, as introduced in Sec.~\ref{sec:noise}, is described by an
Ornstein-Uhlenbeck process in two dimensions. The noise is characterized by eigen-directions,
$\mbf{e}_{i\matM}$, and corresponding relaxation times, $T_i$, of its matrix $\matM$ and by the principal
directions, $\mbf{e}_{i\matV}$, and the corresponding standard deviations, $\sigma_i$, of its covariance matrix $\matV$.
The following values have been chosen (relaxation times are given in units of $\Delta t$).

\begin{subequations}
\begin{eqnarray}
(\mbf{e}_1,\mbf{e}_2)_\matM &=& \begin{pmatrix} 1 & 1 \cr 0 & 2 \end{pmatrix},\quad
                   \mbf{T} \;=\; \begin{pmatrix} 1 \cr 3 \end{pmatrix}\\
(\mbf{e}_1,\mbf{e}_2)_\matV &=& \begin{pmatrix} 1 & -1 \cr 1 & 1 \end{pmatrix},\quad
                   \boldsymbol{\sigma} \;=\; \begin{pmatrix} 0.25 \cr 0.5 \end{pmatrix}
\end{eqnarray}
\end{subequations}

\begin{figure}[h]
\begin{center}
\includegraphics*[width=8.8cm,angle=0]{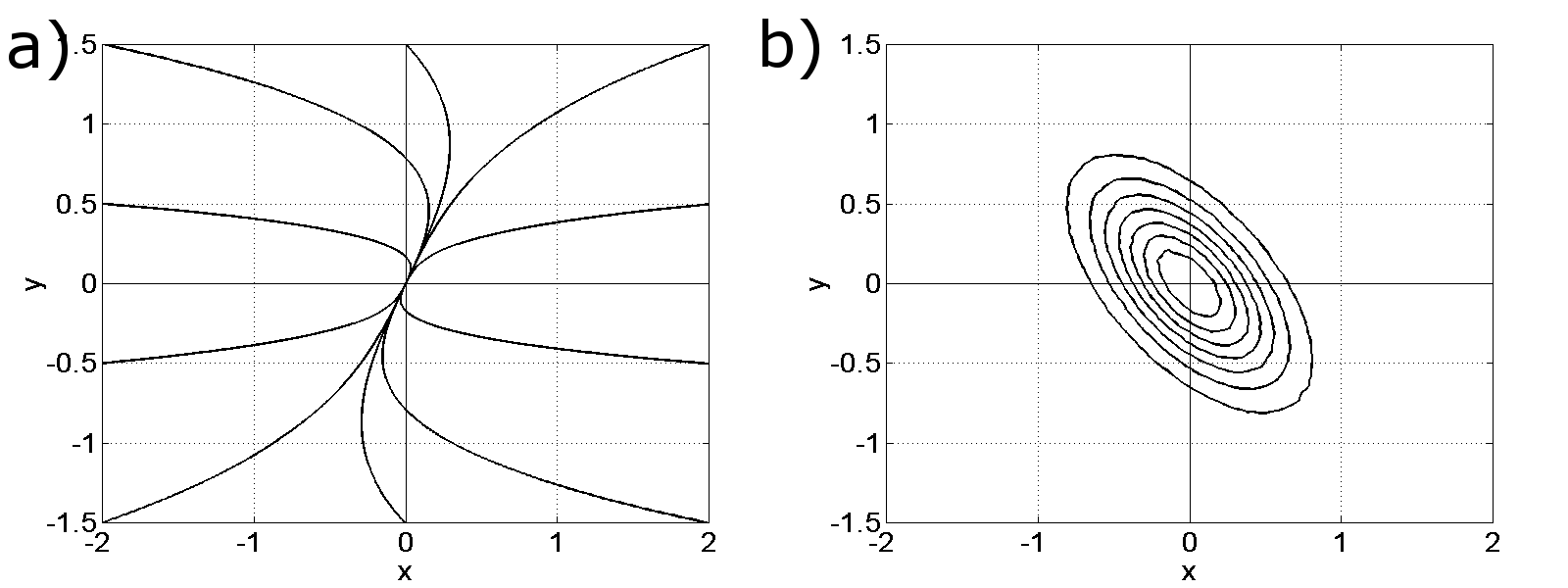}
\end{center}
\caption{\protect Deterministic dynamic and density distribution of the 2D measurement noise $\vecY(t)$.
The trajectories in phase space {\bf (a)} are generated by the deterministic part of the process dynamic
($\dot x=-x+y/3$, $\dot y=-y/3$). There exists a single, attractive, fixed point at the origin.
The contour lines {\bf (b)} of the probability density function have been computed from a time series of $\vecY$.
}
\label{fig_noise}
\end{figure}

\noindent The deterministic part of the dynamic of the measurement noise and the experimental density distribution
of $\vecY$ is visualized in Fig.~\ref{fig_noise}.

\begin{figure}[h]
\begin{center}
\includegraphics*[width=8.8cm,angle=0]{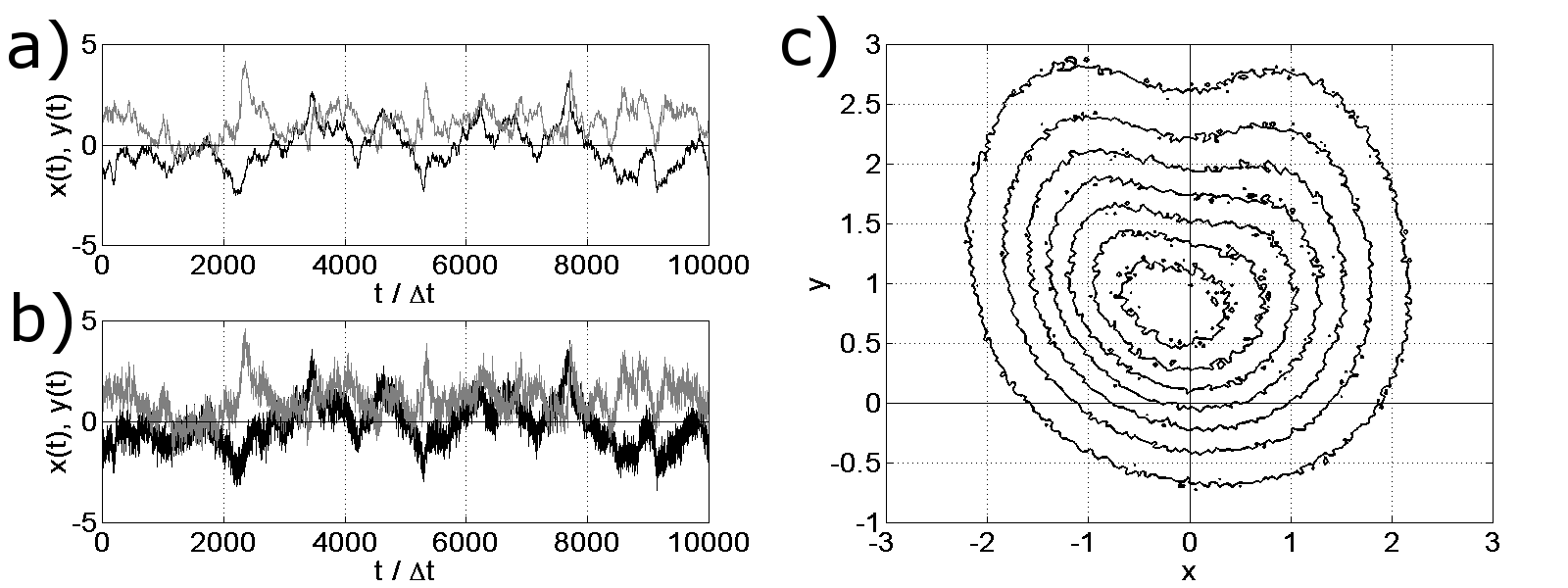}
\end{center}
\caption{\protect Excerpt of a time series of $\vecX(t)$ {\bf (a)} and of a corresponding noisy time series
of $\vecX^*(t)$ {\bf (b)}.
The contour lines {\bf (c)} of the probability density function have been computed from of a noisy time series.
}
\label{fig_noisy}
\end{figure}

\noindent Adding the time series of $\vecX$ and $\vecY$ yields a series of 'noisy' values
$\vecX^*(t)=\vecX(t)+\vecY(t)$. This will be called a noisy time series in the following. Excerpts of $\vecX(t)$
and $\vecX^*(t)$ as well as the experimental density distribution of $\vecX^*$ are shown in Fig.~\ref{fig_noisy}.

\noindent Now the extraction of the measurement noise parameters, as described in
Sec.~\ref{sec:extracting noise}, has been tested. For a sample of 1000 independent realizations of noisy
time series the matrices $\matM(\Delta t)$ and $\matV$ have been estimated. For each estimate a number of
scalar quantities has been calculated. For a characterization of the deterministic part of the noise-dynamic
the angles, $\alpha_i$, spanned by the eigendirections of $\matM$ and the $x$-axis, and the relaxation times,
$T_i$, determined by the eigen-values of $\matM(\Delta t)$, have been used. Their true values are given by

\begin{subequations}
\begin{eqnarray}
\alpha_1 &=& 0^\circ,\quad \alpha_2 \;\approx\; 63.43^\circ\\
T_1/\Delta t &=& 1,\quad T_2/\Delta t \;=\; 3
\end{eqnarray}
\end{subequations}

\noindent The covariance matrix $\matV$ is symmetric and can thus be characterized by three scalars: the angle $\beta$,
spanned by the first principal direction of $\matV$ and the $x$-axis, and the standard deviations $\sigma_i$ in
direction of the principal axes. The true values are given by

\begin{subequations}
\begin{eqnarray}
\beta &=& 45^\circ\\
\sigma_1 &=& 0.25,\quad \sigma_2 \;=\; 0.5
\end{eqnarray}
\end{subequations}

\begin{figure}[h]
\begin{center}
\includegraphics*[width=8.8cm,angle=0]{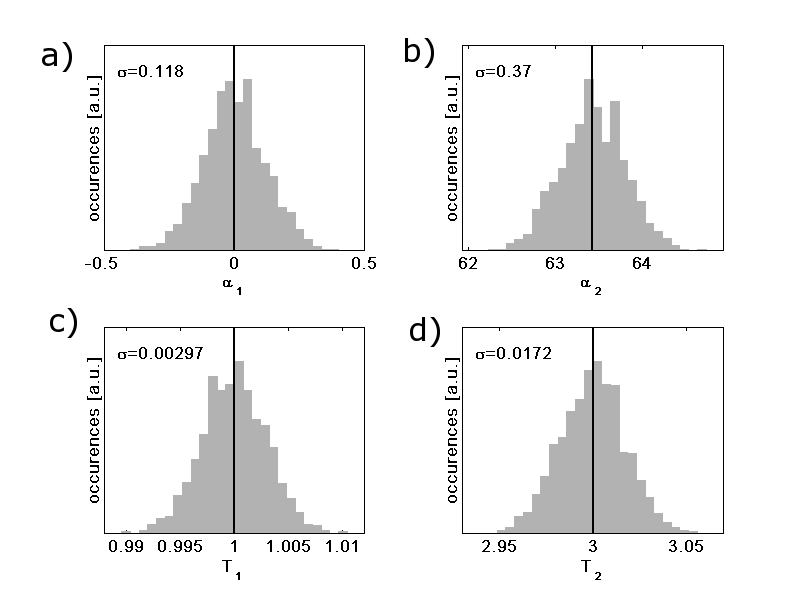}
\end{center}
\caption{\protect Histograms of the observed distributions of the estimated parameters characterizing matrix
$\matM$ of the 2D measurement noise. Angles $\alpha_i$ of the eigen-directions {\bf (a,b)} and corresponding
relaxation times $T_i$ in units of $\Delta t$ {\bf (c,d)}. The standard deviation of the respective
distribution is given by an annotation. The true parameter values are indicated by solid vertical lines.
}
\label{fig_noisefit_M}
\end{figure}

\begin{figure}[h]
\begin{center}
\includegraphics*[width=8.8cm,angle=0]{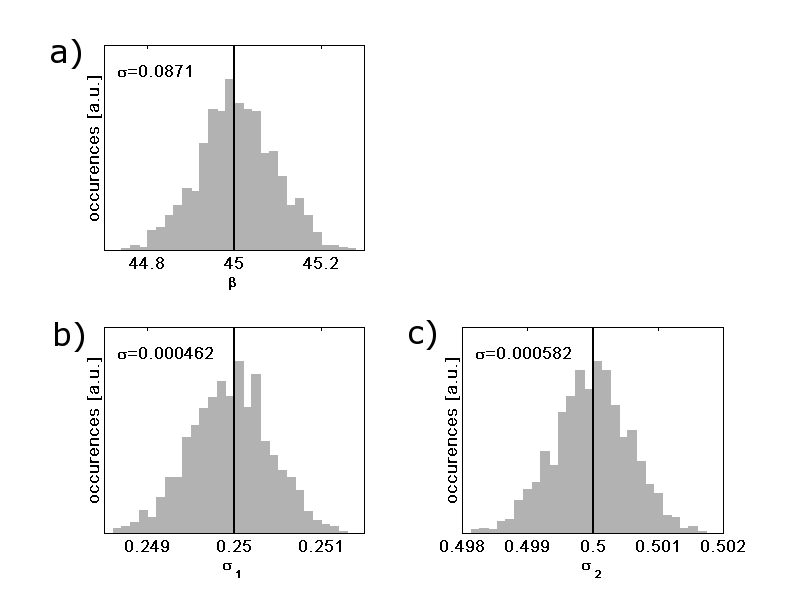}
\end{center}
\caption{\protect Histograms of the observed distributions of the estimated parameters characterizing the
covariance matrix $\matV$ of the 2D measurement noise. Angle $\beta$ of the first principal direction {\bf (a)}
and the principal standard deviations $\sigma_i$ {\bf (b,c)}. The standard deviation of the respective
distribution is given by an annotation. The true parameter values are indicated by solid vertical lines.
}
\label{fig_noisefit_V}
\end{figure}

\noindent Parameter fitting has been performed with a maximum increment $\tau_\text{max}=50\Delta t$ and a maximum
polynomial order of $\nu_\text{max}=3$. The resulting distributions of the estimates are shown in
Figs.~\ref{fig_noisefit_M} and \ref{fig_noisefit_V}. It turns out, that the sample standard deviations of the
angular quantities are given by some tenth of a degree. Relaxation times and noise strengthes are estimated with
relative errors well below one percent.

\noindent For the estimation of the drift- and diffusion functions a complete quadratic ansatz has been made for each component
of $\vecDDrift$ and $\vecDDiff$. Because $\vecDDiff$ is symmetric, this leads to a total of 30 coefficients.
As maximum time increment for the fitting procedure a value of $\tau_\text{max}=15\Delta t$ has been chosen.
The set of weight functions $\Psi$ consisted of 16 Gaussian functions centered at the nodes of a rectangular
$4\times4$ grid covering the $\pm2\sigma$ range of the experimental density distribution of $\vecX^*$.
The standard deviations of the weight functions itself was chosen as twice the distance between neighbouring nodes.

\begin{figure}[h]
\begin{center}
\includegraphics*[width=8.8cm,angle=0]{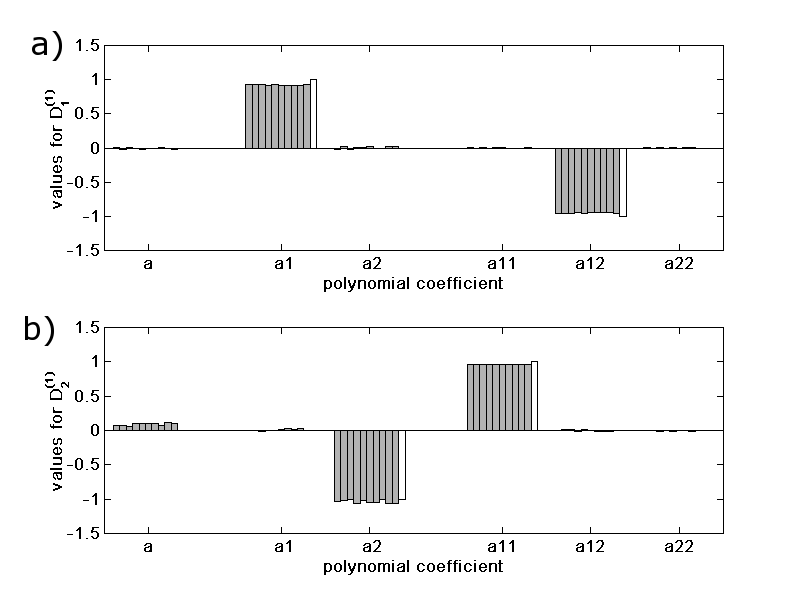}
\end{center}
\caption{\protect Parameter estimates for the drift functions $\DDrift_1$ {\bf (a)} and $\DDrift_2$ {\bf (b)} of
the 2D process, obtained by a linear fit in $\tau$. For each polynomial coefficient the calculated estimates are
given by grey bars with an additional white bar to the right, which shows the true value.
}
\label{fig_D1_linear}
\end{figure}

\begin{figure}[h]
\begin{center}
\includegraphics*[width=8.8cm,angle=0]{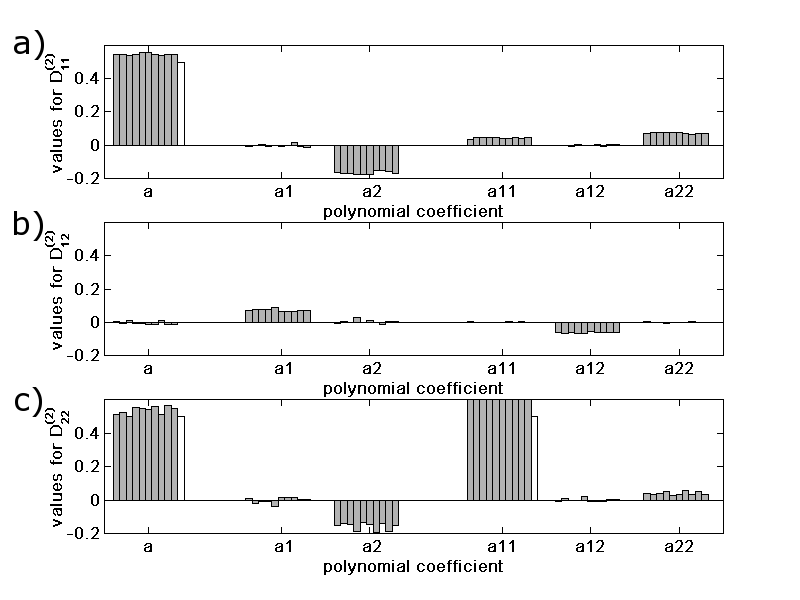}
\end{center}
\caption{\protect Parameter estimates for the diffusion functions $\DDiff_{11}$ {\bf (a)},  $\DDiff_{12}$
{\bf (b)} and $\DDiff_{22}$ {\bf (c)} of the 2D process, obtained by a linear fit in $\tau$. For each polynomial
coefficient the calculated estimates are given by grey bars with an additional white bar to the right, which shows
the true value.
}
\label{fig_D2_linear}
\end{figure}

For this setup the coefficients have been estimated now for a sample of ten independent realizations of the
noisy time series. Using a linear fit in $\tau$ leads to the results shown in Figs.~\ref{fig_D1_linear} and
\ref{fig_D2_linear}.

\begin{figure}[h]
\begin{center}
\includegraphics*[width=8.8cm,angle=0]{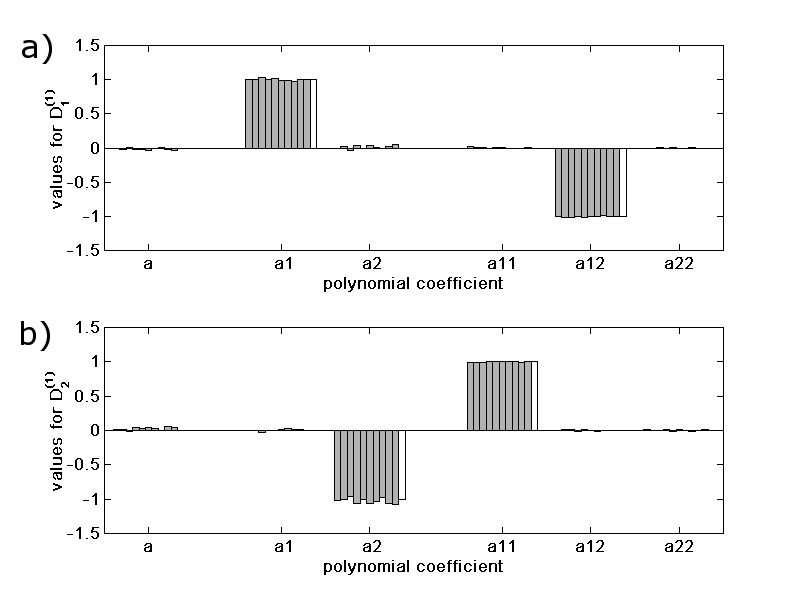}
\end{center}
\caption{\protect Parameter estimates for the drift functions $\DDrift_1$ {\bf (a)} and $\DDrift_2$ {\bf (b)} of
the 2D process, obtained by a quadratic fit in $\tau$. For each polynomial coefficient the calculated estimates are
given by grey bars with an additional white bar to the right, which shows the true value.
}
\label{fig_D1}
\end{figure}

\begin{figure}[h]
\begin{center}
\includegraphics*[width=8.8cm,angle=0]{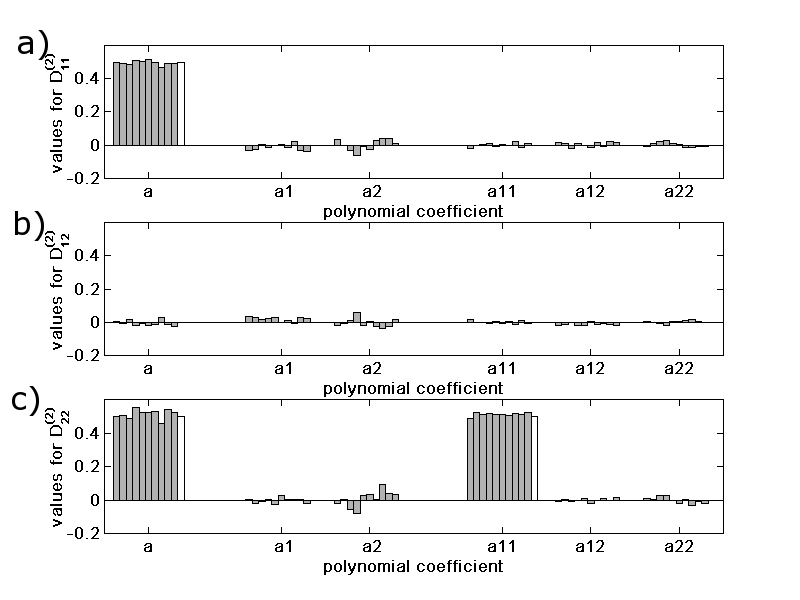}
\end{center}
\caption{\protect Parameter estimates for the diffusion functions $\DDiff_{11}$ {\bf (a)},  $\DDiff_{12}$
{\bf (b)} and $\DDiff_{22}$ {\bf (c)} of the 2D process, obtained by a quadratic fit in $\tau$. For each polynomial
coefficient the calculated estimates are given by grey bars with an additional white bar to the right, which shows
the true value.
}
\label{fig_D2}
\end{figure}

\noindent It can be seen that some of the estimates, especially for the coefficients of the diffusion
functions, are significantly biased. Looking, for example, at coefficient $a_2$ of diffusion function $\DDiff_{11}$
one finds a value of $-0.1663\pm 0.0097$ which significantly differs from the true value of zero. Much better results are
obtained by performing a quadratic fit in $\tau$. To do so, the most simple approach has been chosen: For each parameter
$a_\alpha$ an additional parameter $b_\alpha$ (see Eq.~(\ref{coeff_fitting_final})) is introduced. The only purpose of
this parameters is, to absorb some of the quadratic terms in $\tau$. Because the number of unknowns is doubled this
way, this will also lead to higher fluctuations of the estimates. However, performing such a quadratic fit also
greatly reduces their biasing, as can be seen in Figs.~\ref{fig_D1} and \ref{fig_D2}. For the above mentioned
coefficient $a_2$ of $\DDiff_{11}$, e.g., one now obtaines a value of $0.0025\pm 0.0352$.

\noindent The extraction of noise and process parameters from a noisy time series seems to work for the given 2D test case.
To check if this also holds for higher dimensions, the test case has been extended to four dimensions. Process and
measurement noise now are defined in $x,y,z,w$ space. The process is defined by

\begin{subequations}
\begin{eqnarray}
\vecDDrift(\vecx) &=& \begin{pmatrix} x-xy \cr x^2-y \cr -z \cr -w\end{pmatrix}\\
\vecDDiff(\vecx) &=& \begin{pmatrix} \frac{1}{2} & 0 &0 & 0 \cr 0 & \frac{1\!+\!x^2}{2} & 0 & 0
                                   \cr 0 & 0 & \frac{1\!+\!x^2}{2} & 0 \cr 0 & 0 & 0 & \frac{1\!+\!x^2}{2} \end{pmatrix}
\end{eqnarray}
\end{subequations}

\noindent and the measurement noise by

\begin{subequations}
\begin{eqnarray}
(\mbf{e}_1,\mbf{e}_2,\mbf{e}_3,\mbf{e}_4)_\matM &=& \begin{pmatrix} 1&1&1&0\cr 0&2&0&0\cr 0&0&2&0\cr 0&0&0&1 \end{pmatrix}\\
(\mbf{e}_1,\mbf{e}_2,\mbf{e}_3,\mbf{e}_4)_\matV &=& \begin{pmatrix} 1&\!-1&0&0\cr 1&1&0&0\cr 0&0&1&0\cr 0&0&0&1 \end{pmatrix}
\end{eqnarray}
\end{subequations}

\begin{eqnarray}
\mbf{T} &=& \begin{pmatrix} 1\cr 3\cr 2\cr 2 \end{pmatrix},\quad
\boldsymbol{\sigma} \;=\; \begin{pmatrix} 0.25\cr 0.5\cr 0.25\cr 0.25 \end{pmatrix}.
\end{eqnarray}

\noindent Estimating the relaxation times and the noise strengthes for a noisy time series ($10^7$ points,
$\Delta t=0.005$, $\tau_\text{max}=50$, $\nu_\text{max}=3$) yields the following results.

\begin{eqnarray}
\mbf{\tilde T} &=& \begin{pmatrix} 1.000\cr 3.009\cr 1.997\cr 1.986 \end{pmatrix},\quad
\boldsymbol{\tilde \sigma} \;=\; \begin{pmatrix} 0.2503\cr 0.5008\cr 0.2498\cr 0.2491 \end{pmatrix}
\end{eqnarray}

\noindent The accuracy of the estimated matrices, $\mbf{\tilde M}$ and $\mbf{\tilde V}$, can be expressed in terms of
the relative errors $\boldsymbol{\epsilon}_\matM$ and $\boldsymbol{\epsilon}_\matV$, defined as
$\matM^{-1}\mbf{\tilde M}-\mbf{Id}$ and $\matV^{-1}\mbf{\tilde V}-\mbf{Id}$ respectively. One finds

\begin{subequations}
\begin{eqnarray}
\boldsymbol{\epsilon}_\matM &=& \begin{pmatrix} +0.6&-5.2&-5.2&+2.4\cr
                                                +0.2&+0.8&+2.2&-1.4\cr
                                                +0.5&+0.8&-3.0&+0.8\cr
                                                -1.0&-1.5&+2.0&-1.5 \end{pmatrix}\times 10^{-3}\\
\boldsymbol{\epsilon}_\matV &=& \begin{pmatrix} +0.7&-0.6&+0.7&-1.7\cr
                                                -2.1&+3.1&+1.8&-2.5\cr
                                                -0.8&+3.3&-5.0&+1.9\cr
                                                -0.5&-3.8&+1.9&-1.4 \end{pmatrix}\times 10^{-3}.
\end{eqnarray}
\end{subequations}

\begin{figure}[h]
\begin{center}
\includegraphics*[width=8.8cm,angle=0]{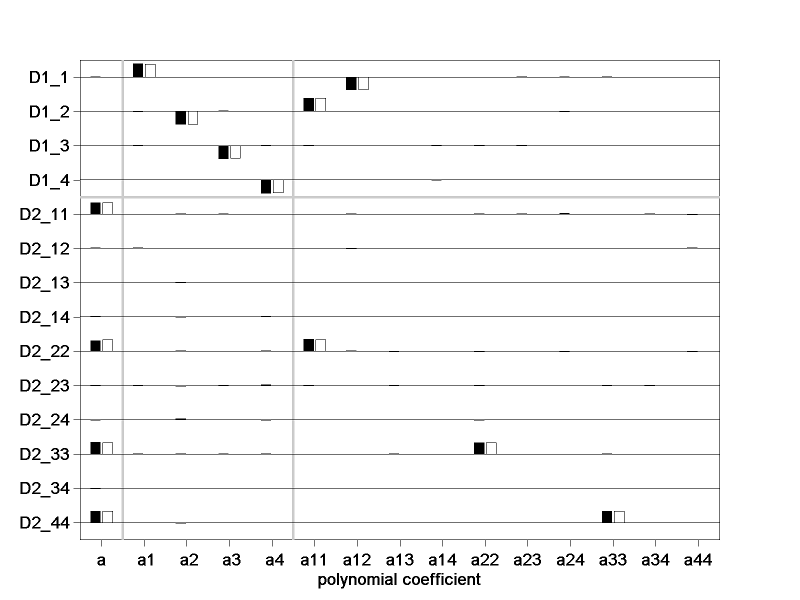}
\end{center}
\caption{\protect Parameter estimates for the polynomial coefficients of the drift- and diffusion functions
of the 4D process, obtained by a quadratic fit in $\tau$. For each coefficient the calculated estimate is
given by a black bar. White bars show the respective true values. Black horizontal lines show the abscissa of the
coordinate systems of the respective Drift- and Diffusion functions denoted to the left.
The results for constant-, linear- and quadratic coefficients as well as the results for drift- and diffusion functions
are seperated by grey lines.
}
\label{fig_D1D2_4d}
\end{figure}

\noindent For the estimation of the drift- and diffusion functions a complete quadratic ansatz has been made
for each component of $\vecDDrift$ and $\vecDDiff$. In four dimensions this leads to a total of 210 coefficients.
As maximum time increment for the fitting procedure a value of $\tau_\text{max}=15\Delta t$ has been chosen.
The set of weight functions $\Psi$ consisted of 81 Gaussian functions centered at the nodes of a rectangular
$3\times3\times3\times3$ grid covering the $\pm2\sigma$ range of the experimental density distribution of $\vecX^*$.
The standard deviations of the weight functions itself was chosen as twice the distance between neighbouring nodes.
Using the same type of quadratic fit in $\tau$ as in the 2D case yields the coefficient estimates
shown in Fig.~\ref{fig_D1D2_4d}.

\section{Conclusions}
\label{sec:conclusions}

A procedure has been described for the analysis of stochastic time series in N dimensions in the presence of strong
measurement noise. The algorithm is able to cope with exponentially correlated noise
and accurately extracts strength and correlation time of the measurement noise as well as the parameters defining the
drift- and diffusion functions of the underlying stochastic process. This has been shown by the analysis of synthetically
generated time series in two and in four dimensions.

The ability to deal with exponentially correlated measurement noise in more than one dimension has not been given by
the approaches available up to now.

Because of the use of weight functions there is no need to perform any density binning. All required quantities can be
obtained from weighted sums of the values of the time series. This avoids the aliasing errors caused by finite
bin sizes.

All calculation have been performed on a standard desktop PC. The analysis of a signal took about five minutes (2D case)
respectively fifty minutes (4D case). 

In the current implementation only a simplified quadratic fit in the increments $\tau$ can be performed. Implementing
a full polynomial fit, as mentioned in Sec.~\ref{sec:extracting coeffs}, should allow to extend the range of time
increments that can be used for the analysis and thus should increase the accuracy of the results.
This has to be done in the future.

Another point to be improved is the restriction on polynomial approximations of the drift- and diffusion functions. An
approximation by spline-based functions would be much more flexible.
When using such a parametrization, however, it will no longer be possible to accurately express the convolutions in
Eq.~(\ref{weighted_m1m2}) in terms of observable quantities. It will become neccessary to also introduce a parametrization
for the density $\mom0$ which significantly complicates the calculations and also introduces additional parameters
to be estimated.

Also a future task is the application to some real world data.

\section{Acknowledgements}

The author especially wants to thank Joachim Peinke, Rudolf Friedrich, Maria Haase, David Kleinhans and Pedro G. Lind
for useful discussions.

\appendix
\section{Gauss functions}
\label{app:gauss}
%
\subsection{Index-vectors and monomials}
\label{app:gauss:monomials}

For the sake of a compact syntax, multiple indices will frequently be combined into an
index-vector. For example $A_{j_1\ldots j_n}$ will be written as $A_\mbf{j}$. To denote the length of
such a vector $\mbf j$, the function $\ell(\mbf{j})$ will be used.

As a further abbreviation the symbol $\Mon$ is introduced for monomials of the components of a vector.
A monomial $x_{j_1}\ldots x_{j_n}$ will be written as $\Mon_{j_1\ldots j_n}(\vecx)$ or simply as
$\Mon_\mbf{j}(\vecx)$.
Monomials of the nabla vector will be used, to denote multiple partial differentiation more compactly
by $\Mon_\mbf{j}(\nabla)$.

%
\subsection{Fourier transform and convolution}
\label{app:gauss:basic}

Let Fourier transform and convolution of functions $\mathbb{R}^N\to\mathbb{C}$ be defined as below.
For notational simplicity the `hat' syntax will be used to denote the Fourier transform of single functions.
For more complex expressions the functional form $\Four(\ldots)$ will usually be the better choice.

\begin{eqnarray}
\Four[f(\vecx)](\vecomega) &=& \hat f(\vecomega)
                          \;:=\; \int_\vecx e^{-i \vecomega^t\vecx} f(\vecx)\bdif x    \label{def_fourier}\\
f(\vecx)*g(\vecx) &:=& \int_{\vecx'}f(\vecx')g(\vecx-\vecx')\bdif x'\cr
                  &=&  \int_{\vecx'}f(\vecx-\vecx')g(\vecx')\bdif x'                   \label{def_conv}
\end{eqnarray}

\noindent Above definitions imply the following properties.

\begin{subequations}
\begin{eqnarray}
\Four[\fracpp{x_j}f(\vecx)] &=& i\omega_j \hat f(\vecomega)                     \label{fourier_3a}\\
\Four[x_jf(\vecx)] &=& i\fracpp{\omega_j} \hat f(\vecomega)                     \label{fourier_3b}\\[0.2em]
\Four[f(\vecx)*g(\vecx)] &=& \hat f(\vecomega)\hat g(\vecomega)                 \label{fourier_3c}\\[0.5em]
\fracpp{x_j}[f(\vecx)*g(\vecx)] &=& [\fracpp{x_j}f(\vecx)]*g(\vecx)\cr
                                &=& f(\vecx)*[\fracpp{x_j}g(\vecx)]             \label{fourier_3d}
\end{eqnarray}\label{basic:prop3}
\end{subequations}

%
\subsection{Derivatives and monomial products of Gauss functions}
\label{app:gauss:deriv}

Let $G(\matC,\vecx)$ denote a normalized Gauss function with covariance matrix $\matC$ and
function argument $\vecx\in\mathbb{R}^N$.

\begin{eqnarray}
G(\matC,\vecx) &:=& \frac{1}{\sqrt{(2\pi)^N|\text{det}(\matC)|}}\text{e}^{-\frac{1}{2}\vecx^t\matC^{-1}\vecx}
\label{normGauss}
\end{eqnarray}

\noindent This function is a eigenfunction of the Fourier transform.

\begin{eqnarray}
\hat G(\matC,\vecomega) &=& \text{e}^{-\frac{1}{2}\vecomega^t\matC\vecomega}\cr
                        &=& \sqrt{(2\pi)^N|\text{det}(\matC^{-1})|}\;G(\matC^{-1},\vecomega)
                                                                                     \label{fourier_gauss}
\end{eqnarray}

\noindent It can be shown by mathematical induction, that the derivatives of $G$ all have the form

\begin{eqnarray}
\Mon_\mbf{j}(\nabla_x)G(\matC,\vecx) &=& P_\mbf{j}(\matC,\vecx)G(\matC,\vecx),      \label{deriv_gauss}
\end{eqnarray}

\noindent where $P$ is a polynomial of order $\ell(\mbf{j})$ in $\vecx$. Mathematical induction also shows,
that $P$ only contains monomials in $\vecx$ of either even or odd order. The coefficients of $P$ can be
expressed in terms of the elements of $\matC^{-1}$, but no attempt will be made here to give an
explicit formula, because the expressions for any finite order polynomial can be derived iteratively. Up
to order three the derivatives of $G$ are given by (using summation convention)

\begin{subequations}
\begin{eqnarray}
\fracpp{x_j}G
    &=& \Big[ -C^{-\!1}_{j\alpha}\,x_{\alpha} \Big]\,G\\
\frac{\p^2}{\p x_j\p x_k}G
    &=& \Big[ C^{-\!1}_{j\alpha}C^{-\!1}_{k\beta}\,x_{\alpha}x_{\beta}
              -C^{-\!1}_{jk} \Big]\,G\\
\frac{\p^3}{\p x_j\p x_k\p x_l}G
    &=& \Big[ -C^{-\!1}_{j\alpha}C^{-\!1}_{k\beta}C^{-\!1}_{l\gamma}\,x_\alpha x_\beta x_\gamma \cr
     &&     +\big(C^{-\!1}_{jk}C^{-\!1}_{l\alpha} +C^{-\!1}_{jl}C^{-\!1}_{k\alpha} \phantom{\Big[}\cr
     &&   \;\;\;\;       +C^{-\!1}_{kl}C^{-\!1}_{j\alpha}\big)\,x_\alpha \Big]\,G.
\end{eqnarray}
\end{subequations}

\noindent Terms of the form $\Mon(\vecx) G$ will be called monomial products of $G$ in the following.
Such products can be expressed in terms of derivatives of $G$. Applying a Fourier transform to
Eq.~(\ref{deriv_gauss}) and using of Eqs.~(\ref{fourier_3a}), (\ref{fourier_3b}) and (\ref{fourier_gauss})
first gives

\begin{eqnarray}
\Mon_\mbf{j}(i\vecomega)G(\matC^{-1},\vecomega)
    &=& P_\mbf{j}(\matC,i\nabla_\omega)G(\matC^{-1},\vecomega).
\end{eqnarray}

\noindent Substituting $i\vecomega$ by $\vecx$ and $\matC$ by $-\matC^{-1}$ (thus
$\vecomega^t\matC\vecomega$ by $\vecx^t\matC^{-1}\vecx$) then finally yields

\begin{eqnarray}
\Mon_\mbf{j}(\vecx)G(\matC,\vecx) &=& P_\mbf{j}(-\matC^{-1},-\nabla_x)G(\matC,\vecx).   \label{monprod_gauss}
\end{eqnarray}

\noindent Up to order three the monomial products of $G$ therefore read

\begin{subequations}
\begin{eqnarray}
x_jG
    &=& \Big[ -C_{j\alpha}\,\fracpp{x_\alpha} \Big]\,G\\
x_jx_kG
    &=& \Big[ C_{j\alpha}C_{k\beta}\;\frac{\p^2}{\p x_\alpha\p x_\beta}
              +C_{jk} \Big]\,G\\
x_jx_kx_lG
    &=& \Big[ -C_{j\alpha}C_{k\beta}C_{l\gamma}\,\frac{\p^3}{\p x_\alpha\p x_\beta\p x_\gamma} \cr
     &&     -\big(C_{jk}C_{l\alpha} +C_{jl}C_{k\alpha} \phantom{\Big[}\cr
     &&   \;\;\;\;       +C_{kl}C_{j\alpha}\big)\,\fracpp{x_\alpha} \Big]\,G.
\end{eqnarray}
\end{subequations}

%
\subsection{Moments of Gauss functions}
\label{app:gauss:moments}

Integrating Eq.~(\ref{monprod_gauss}) with respect to $\vecx$ yields expressions for the moments
of $G$. Because integrals of derivatives of $G$ are vanishing, the moment is determined by the
constant part of the polynomial $P_\mbf{j}$. This coefficient will be non-zero only for even moments. The odd
moments of $G$ all evaluate to zero (as can also bee seen from symmetry considerations). The first
non-vanishing moments are given by

\begin{subequations}
\begin{eqnarray}
\int_\vecx G\bdif x
    &=& 1\\
\int_\vecx x_jx_kG\bdif x
    &=& C_{jk}\\
\int_\vecx x_jx_kx_lx_mG\bdif x
    &=& C_{jk}C_{lm} +C_{jl}C_{km}\cr
    && +C_{jm}C_{kl}.
\end{eqnarray}
\end{subequations}

%
\subsection{Gauss functions in convolutions}
\label{app:gauss:conv}

Convolutions of the form $[\Mon_\mbf{j}(\vecx)G(\matC,\vecx)]\!*\!\!f(\vecx)$ can be expressed in terms of
derivatives of the convolution $G*f$. This can be derived straightforwardly by first expressing
$\Mon(\vecx)G$ by derivatives of $G$ and then applying Eq.~(\ref{fourier_3d}). One finds

\begin{eqnarray}
[\Mon_\mbf{j}(\vecx)G(\matC,\vecx)]\!*\!\!f(\vecx)
  &=& P_\mbf{j}(-\matC^{-1}\!\!,-\nabla_x)\cr
     &&\times [G(\matC,\vecx)\!*\!\!f(\vecx)].
\end{eqnarray}

\noindent It is also possible to express convolutions of the form
$G(\matC,\vecx)\!*\![\Mon_\mbf{j}(\vecx)f(\vecx)]$ by derivatives of monomial products of $G*f$. This can be
derived in Fourier space. So let $F$ denote the Fourier transform of the expression under consideration.

\begin{eqnarray}
F &:=& \Four\Bigl\{G(\matC,\vecx)\!*\!\big[\Mon_\mbf{j}(\vecx)f(\vecx)\big]\Bigr\}\cr
  &=& \hat G(\matC,\vecomega)\Mon_\mbf{j}(i\nabla_\omega)\hat f(\vecomega)
\end{eqnarray}

\noindent Using the identity

\begin{eqnarray}
\frac{1}{\hat G(\matC,\vecomega)}
  &=& \sqrt{(2\pi)^N|\text{det}(\matC^{-1})|}\;G(\matC^{-1}\!,i\vecomega)     \label{gauss_inv}
\end{eqnarray}

\noindent leads to

\begin{eqnarray}
F &=& \hat G(\matC,\vecomega)\Mon_\mbf{j}(i\nabla_\omega)
         \big[\frac{\hat G(\matC,\vecomega)}{\hat G(\matC,\vecomega)}\hat f(\vecomega)\big]\cr
 &=& \hat G(\matC,\vecomega)\sqrt{(2\pi)^N|\text{det}(\matC^{-1})|}
       \;\Mon_\mbf{j}(i\nabla_\omega)\cr
  && \times\Big\{G(\matC^{-1}\!,i\vecomega)\big[\hat G(\matC,\vecomega)\hat f(\vecomega)\big]\Big\}.
\end{eqnarray}

\noindent Now the product rule of differentiation is applied to the term in the curly brackets. Using
index-vectors the product rule can be written as

\begin{eqnarray}
\Mon_\mbf{j}(\nabla)[f(\vecx)g(\vecx)]
  &=& \sum_{(\mbf{j}',\mbf{j}'')\in {\cal P}(\mbf{j})}
      \big[\Mon_\mbf{j'}(\nabla)f(\vecx)\big]\cr
   && \times\big[\Mon_\mbf{j''}(\nabla)g(\vecx)\big].
\end{eqnarray}

\noindent Here ${\cal P}(\mbf{j})$ denotes the set of all $2^{\ell(\mbf{j})}$ pairs
$(\mbf{j}',\mbf{j}'')$ that can be obtained by distributing the components of $\mbf{j}$ on two vectors
$\mbf{j'}$ and $\mbf{j''}$. Applying the product rule yields

\begin{widetext}

\begin{eqnarray}
F &=& \hat G(\matC,\vecomega)\sqrt{(2\pi)^N|\text{det}(\matC^{-1})|}
      \sum_{(\mbf{j}',\mbf{j}'')\in {\cal P}(\mbf{j})}
      \Big\{\Mon_\mbf{j'}(i\nabla_\omega)\; G(\matC^{-1}\!,i\vecomega)\Big\}
      \Big\{\Mon_\mbf{j''}(i\nabla_\omega)\big[\hat G(\matC,\vecomega)\hat f(\vecomega)\big]\Big\}.
\end{eqnarray}

\noindent Temporarily substituting $\mbf{z}=-i\vecomega$ in the first bracket and using
$G(\bullet,-\vecx)=G(\bullet,\vecx)$ gives

\begin{eqnarray}
\Mon_\mbf{j'}(i\nabla_\omega)\;G(\matC^{-1}\!,i\vecomega)
  &=& \Mon_\mbf{j'}(\nabla_z)\; G(\matC^{-1}\!,z)
  \;=\; P_\mbf{j'}(\matC^{-1}\!,z)\; G(\matC^{-1}\!,z)
  \;=\; P_\mbf{j'}(\matC^{-1}\!,-i\vecomega)\; G(\matC^{-1}\!,i\vecomega).
\end{eqnarray}

\noindent Now $G(\matC^{-1}\!,i\vecomega)$ can be written in front of the sum. Using
Eq.~(\ref{gauss_inv}) some factors cancel out and it remains

\begin{eqnarray}
F &=& \sum_{(\mbf{j}',\mbf{j}'')\in {\cal P}(\mbf{j})}
      P_\mbf{j'}(\matC^{-1}\!,-i\vecomega)
      \Mon_\mbf{j''}(i\nabla_\omega)\big[\hat G(\matC,\vecomega)\hat f(\vecomega)\big].
\end{eqnarray}

\noindent Switching back to real space finally gives the desired relation

\begin{eqnarray}
G(\matC,\vecx)*\big[\Mon_\mbf{j}(\vecx)f(\vecx)\big]
  &=& \sum_{(\mbf{j}',\mbf{j}'')\in {\cal P}(\mbf{j})}
      P_\mbf{j'}(\matC^{-1}\!,-\nabla_x)
      \Big\{\Mon_\mbf{j''}(\vecx)\big[G(\matC,\vecx)*f(\vecx)\big]\Big\}.
\end{eqnarray}

\section{Conditioned moments $\bmom{k}$}
\label{app:cond_mom}

The somewhat lengthy calculations leading to Eq.~(\ref{noisy_moments}) are given below. The function argument of
$\matM(\tau)$and  $\matC(\tau)$ will be omitted for notational simplicity. Partial derivation with respect to $x_i$
will be denoted by $\p_i$  and Einsteins summation convention will be used.
Starting with Eq.~(\ref{def_condmom_noisy}), inserting Eqs.~(\ref{pdf_noise}) and (\ref{pdf_noisy}) and interchanging
the order of integration gives

\begin{subequations}
\begin{eqnarray}
\moms{0}(\vecx) &=&\int_\vecz G(\matV,\vecx\!-\!\vecz)\int_{\vecz'}\!\rho(\vecz,\vecz',\tau)\int_{\vecx'}
  G(\matC,\vecx'\!-\!\vecz'\!-\!\matM\!\cdot\!(\vecx\!-\!\vecz))\bdif x'\bdif z'\bdif z\\
\moms{1}_i(\vecx,\tau) &=&\int_\vecz G(\matV,\vecx\!-\!\vecz)\int_{\vecz'}\!\rho(\vecz,\vecz',\tau)\int_{\vecx'}
  (x'_i\!-\!x_i)\,G(\matC,\vecx'\!-\!\vecz'\!-\!\matM\!\cdot\!(\vecx\!-\!\vecz))\bdif x'\bdif z'\bdif z\\
\moms{2}_{ij}(\vecx,\tau) &=&\int_\vecz G(\matV,\vecx\!-\!\vecz)\int_{\vecz'}\!\rho(\vecz,\vecz',\tau)\int_{\vecx'}
  (x'_i\!-\!x_i)(x'_j\!-\!x_j)\,G(\matC,\vecx'\!-\!\vecz'\!-\!\matM\!\cdot\!(\vecx\!-\!\vecz))\bdif x'\bdif z'\bdif z
\end{eqnarray}
\end{subequations}

\noindent The integrals with respect to $\vecx'\!$ can be expressed in terms of the moments of the involved Gauss
function (see Sec.~(\ref{app:gauss:moments})). Using the definition of $\mom{0}$ then allows to express $\moms{0}$ by
a convolution.

\begin{eqnarray}
\moms{0}(\vecx) &=& G(\matV,\vecx)*\mom{0}(\vecx)\label{def_moms1} \\
\end{eqnarray}

\noindent The other moments so far read

\begin{subequations}
\begin{eqnarray}
\moms{1}_i(\vecx,\tau) &=&\int_\vecz G(\matV,\vecx\!-\!\vecz)\int_{\vecz'}\!\rho(\vecz,\vecz',\tau)
\left[z'_i\!-\!x_i\!+\!M_{ii'}(x_{i'}\!-\!z_{i'}))\right]\bdif z'\bdif z\\
\moms{2}_{ij}(\vecx,\tau) &=&\int_\vecz G(\matV,\vecx\!-\!\vecz)\int_{\vecz'}\!\rho(\vecz,\vecz',\tau)
       [C_{ij}+(z'_i\!-\!x_i\!+\!M_{ii'}(x_{i'}\!-\!z_{i'})) \cr
     &&\times       (z'_j\!-\!x_j\!+\!M_{jj'}(x_{j'}\!-\!z_{j'})) ]\bdif z'\bdif z.
\end{eqnarray}
\end{subequations}

\noindent Sorting the terms in rectangular brackets by powers of the components of $\vecz'\!-\vecz$ and using the
definitions of the moments $\bmom{k}$, allows the intergals with respect to $\vecz'\!$ to be expressed by the
moments $\bmom{k}$.

\begin{subequations}
\begin{eqnarray}
\moms{1}_i(\vecx,\tau) &=&\int_\vecz G(\matV,\vecx\!-\!\vecz)
\left[\mom{1}_i(\vecz,\tau)\!-\!(\delta_{ii'}\!-\!M_{ii'})(x_{i'}\!-\!z_{i'})\,\mom{0}(\vecz)\right]\bdif z\\
\moms{2}_{ij}(\vecx,\tau) &=&\int_\vecz G(\matV,\vecx\!-\!\vecz)
       \left[\mom{2}_{ij}(\vecz,\tau)\!+\!C_{ij}\mom{0}(\vecz)-(\delta_{jj'}\!-\!M_{jj'})(x_{j'}\!-\!z_{j'}\,)
                \mom{1}_i(\vecz,\tau)\right.\cr
       && -(\delta_{ii'}\!-\!M_{ii'})(x_{i'}\!-\!z_{i'})\,\mom{1}_j(\vecz,\tau)
         \left.+(\delta_{ii'}\!-\!M_{ii'})(x_{i'}
               \!-\!z_{i'})(\delta_{jj'}\!-\!M_{jj'})(x_{j'}\!-\!z_{j'})\,\mom{0}(\vecz)\right]\bdif z.
\end{eqnarray}
\end{subequations}

\noindent Now the relation $\int_\vecz (x_i-z_i)f(\vecx-\vecz)g(\vecz)\bdif z=[x_if(\vecx)]*g(\vecx)$ can be used to
express the noisy moments as convolutions. Function arguments can now be omitted without confusion ($G$ refers to
$G(\matV,\vecx)$).

\begin{subequations}
\begin{eqnarray}
\moms{1}_i &=&G*\mom{1}_i-(\delta_{ii'}\!-\!M_{ii'})[x_{i'}G]*\mom{0}\\
\moms{2}_{ij} &=& G*\mom{2}_{ij} +C_{ij}\,G*\mom{0}
           -(\delta_{ii'}\!-\!M_{ii'})[x_{i'}G]*\mom{1}_j
           -(\delta_{jj'}\!-\!M_{jj'})[x_{j'}G]*\mom{1}_i\cr
         &&+(\delta_{ii'}\!-\!M_{ii'})(\delta_{jj'}\!-\!M_{jj'})[x_{i'}x_{j'}G]*\mom{0}.
\end{eqnarray}
\end{subequations}

\noindent Because $G$ is a Gauss function, the monomial products $x_iG$ and $x_ix_jG$ can be expressed by
derivatives. Inserting the definition of $\matC$ (Eq.~(\ref{def_matrixC})) and resorting terms subsequently leads to

\begin{subequations}
\begin{eqnarray}
\moms{1}_i &=&G*\mom{1}_i+(\delta_{ii'}\!-\!M_{ii'})V_{i'k}[\p_{k}G]*\mom{0}\\
\moms{2}_{ij} 
        &=& G*\mom{2}_{ij} +\{(\delta_{jj'}\!-\!M_{jj'})V_{ij'}+(\delta_{ii'}\!-\!M_{ii'})V_{i'j}\}G*\mom{0}
           +(\delta_{ii'}\!-\!M_{ii'})V_{i'k}[\p_{k}G]*\mom{1}_j\cr
         &&+(\delta_{jj'}\!-\!M_{jj'})V_{j'k}[\p_{k}G]*\mom{1}_i
           +(\delta_{ii'}\!-\!M_{ii'})V_{i'k}(\delta_{jj'}\!-\!M_{jj'})V_{j'l}[\p_{k}\p_{l}G]*\mom{0}.
\end{eqnarray}
\end{subequations}

\noindent Introducing the abbreviation $\matQ:=(\mbf{Id}-\matM)\matV$ and using the relation $(\p f)*g=\p(f*g)$
this can be written as

\begin{subequations}
\begin{eqnarray}
\moms{1}_i &=&G*\mom{1}_i+Q_{ii'}\p_{i'}(G*\mom{0})\\
\moms{2}_{ij} &=& G*\mom{2}_{ij}+(Q_{ij}\!+Q_{ji})\,G*\mom{0}
           +Q_{ii'}\p_{i'}(G*\mom{1}_j)
           +Q_{jj'}\p_{j'}(G*\mom{1}_i)\cr
         &&+Q_{ii'}Q_{jj'}\,\p_{i'}\p_{j'}(G*\mom{0}).
\end{eqnarray}
\end{subequations}

\noindent Substituting $G*\mom{0}=\moms{0}$ according to Eq.~(\ref{def_moms1}) gives

\begin{subequations}
\begin{eqnarray}
\moms{1}_i &=&G*\mom{1}_i + Q_{ii'}\p_{i'}\moms{0}\\
\moms{2}_{ij} &=&G*\mom{2}_{ij}  + (Q_{ij}+Q_{ji})\,\moms{0}
           +Q_{ii'}\p_{i'}(G*\mom{1}_j)
           +Q_{jj'}\p_{j'}(G*\mom{1}_i)
           +Q_{ii'}Q_{jj'}\,\p_{i'}\p_{j'}\moms{0}.
\end{eqnarray}
\end{subequations}

\noindent Now $G*\mom{1}_i=\moms{1}_i-Q_{ii'}\p_{i'}\moms{0}$ can be substituted, what leads to

\begin{eqnarray}
\moms{2}_{ij} &=&G*\mom{2}_{ij}  + (Q_{ij}+Q_{ji}-Q_{ii'}Q_{jj'}\,\p_{i'}\p_{j'})\,\moms{0}
           +Q_{ii'}\p_{i'}\moms{1}_j+Q_{jj'}\p_{j'}\moms{1}_i.
\end{eqnarray}

\noindent Putting these results together and using Eq.~(\ref{def_moments}) to express $\bmom{k}$ in terms of $\bh{k}$
and $\mom{0}$ yields the final expressions for the noisy moments $\bmoms{k}$.

\begin{subequations}
\begin{eqnarray}
\moms{0} &=&G*\mom{0}\\
\moms{1}_i &=&G*(\h{1}_i\mom{0}) + Q_{ii'}\p_{i'}\moms{0}\\
\moms{2}_{ij} &=&G*(\h{2}_{ij}\mom{0})  + (Q_{ij}+Q_{ji}-Q_{ii'}Q_{jj'}\,\p_{i'}\p_{j'})\,\moms{0}
           +Q_{ii'}\p_{i'}\moms{1}_j+Q_{jj'}\p_{j'}\moms{1}_i
\end{eqnarray}
\end{subequations}

\end{widetext}


%

\begin{thebibliography}{00}

\bibitem{friedrich11} R.~Friedrich, J.~Peinke, M.~Sahimi, and T.M.R.~Rahimi,
                      Phys.~Rep.~{\bf 506}, 87 (2011)

\bibitem{friedrich08} R.~Friedrich, J.~Peinke, and M.R.R.~Tabar,
                      {\it Complexity in the view of stochastic processes}
                      in Springer Encyclopedia of Complexity and Systems Science
                      (Springer, Berlin, 2008)

\bibitem{schreiberbook} H.~Kantz and T.~Schreiber, 
                        {\it Nonlinear Time Series Analysis},
                        (Cambridge University Press, Cambridge, England, 1997)

\bibitem{abarbanel} H.~D.~I.~Abrabanel, R.~Brown, J.~J.~Sidorowich, and L.~S.~Tsimiring, 
                        Rev.~Mod.~Phys.~{\bf 65}, 1331 (1993)

\bibitem{friedrich97} R.~Friedrich and J.~Peinke,
                      Phys.~Rev.~Lett.~{\bf 78}, 863 (1997)

\bibitem{ryskin97} G.~Ryskin,
                      Phys.~Rev.~E~{\bf 56}, 5123 (1997)

\bibitem{siegert98} S.~Siegert,~R.~Friedrich, and J.~Peinke,
                      Phys.~Lett.~A~{\bf 243}, 275 (1998)

\bibitem{friedrich00} R.~Friedrich et al.,
                      Phys.~Lett.~A~{\bf 271}, 217 (2000)

\bibitem{gradisek00} J.~Gradisek, S.~Siegert, R.~Friedrich, and I.~Grabec,
                      Phys.~Rev.~E~{\bf 62}, 3146 (2000)

\bibitem{finance00} R.~Friedrich, J.~Peinke, and Ch.~Renner,
                      Phys.~Rev.~Lett.~{\bf 84}, 5224 (2000)

\bibitem{trafic02} S.~Kriso, J.~Peinke, R.~Friedrich, and P.~Wagner,
                      Phys.~Lett.~A~{\bf 299}, 287 (2002)

\bibitem{circuit03} M.~Siefert, A.~Kittel, R.~Friedrich, and J.~Peinke,
                      Europhys.~Lett.~{\bf 61}, 466 (2003)

\bibitem{circuit04} M.~Siefert and J.~Peinke,
                      Int.~J.~Bifurcation~Chaos~Appl.~Sci.~Eng.~{\bf 14}, 2005 (2004)

\bibitem{heart04} T.~Kuusela,
                      Phys.~Rev.~E~{\bf 69}, 031916 (2004)

\bibitem{climate04} C.~Collette and M.~Ausloos,
                      Int.~J.~Mod.~Phys.~C~{\bf 15}, 1353 (2004)

\bibitem{climate05} P.G.~Lind, A.~Mora, J.A.C.~Gallas and M.~Haase,
                      Phys.~Rev.~E~{\bf 72}, 056706 (2005)

\bibitem{fluid07} A.P.~Nawroth, J.~Peinke, D.~Kleinhans, and R.~Friedrich,
                      Phys.~Rev.~E~{\bf 76}, 056102 (2007)

\bibitem{epilepsy08} J.~Prusseit and K.~Lehnertz,
                      Phys.~Rev.~E~{\bf 77}, 041914 (2008)

\bibitem{epilepsy09} D.~Lamouroux, and K.~Lehnertz,
                      Phys.~Lett.~A~{\bf 373}, 3507 (2009)


\bibitem{sampling02} P.~Sura and J.~Barsugli,
                      Phys.~Lett.~A~{\bf 305}, 304 (2002)


\bibitem{sampling05} D.~Kleinhans, R.~Friedrich, A.~Nawroth, and J.~Peinke,
                      Phys.~Lett.~A~{\bf 346}, 42 (2005)

\bibitem{honisch11} C.~Honisch and R.~Friedrich,
                      Phys.~Rev.~E~{\bf 83}, 066701 (2011)

\bibitem{sampling12MLE} D.~Kleinhans,
                      Phys.~Rev.~E~{\bf 85}, 026705 (2012)

\bibitem{sampling08} J.~Gottschall and J.~Peinke,
                      New~J.~Phys.~{\bf 10}, 083034 (2008)

\bibitem{noise93} E.~J.~Kostelich and T.~Schreiber,
                      Phys.~Rev.~E~{\bf 48}, 1752 (1993)

\bibitem{noise00} J.~P.~M.~Heald and J.~Stark,
                      Phys.~Rev.~Lett.~{\bf 84}, 2366 (2000)

\bibitem{iterative06} F.~Boettcher, J.~Peinke, D.~Kleinhans, R.~Friedrich, P.G.~Lind, and M.~Haase,
                      Phys.~Rev.~Lett.~{\bf 97}, 090603 (2006)

\bibitem{iterative10} P.G.~Lind, M.~Haase, F.~Boettcher, J.~Peinke, D.~Kleinhans, R.~Friedrich,
                      Phys.~Rev.~E~{\bf 81}, 041125 (2010)

\bibitem{lehle11} B.~Lehle,
                      Phys.~Rev.~E~{\bf 83}, 021113 (2011)

\bibitem{platen99} P.~E.~Kloeden and E.~Platen,
                      {\it Numerical Solution of Stochastic Differential Equations}
                      (Springer, New York, 1999)

\bibitem{risken89} H.~Risken,
                      {\it The Fokker-Planck Equation}
                      (Springer, New York, 1989)

\end{thebibliography}
\end{document}